%% file: sn-article-main.tex
\documentclass[sn-standardnature]{sn-jnl}
\usepackage{physics}
\jyear{2023}%

\theoremstyle{thmstyleone}%
\theoremstyle{thmstyletwo}%
\theoremstyle{thmstylethree}%

\begin{document}

\title[Quantum simulation of actinide chemistry with quantum computers]{Quantum simulation of actinide chemistry: towards scalable algorithms on trapped ion quantum computers}

\author*[1]{\fnm{Kesha} \sur{Sorathia}}\email{kesha.sorathia@quantinuum.com}
\author[1] {\fnm{Cono} \sur{Di Paola}}\email{}

\author[1]{\fnm{Gabriel} \sur{Greene-Diniz}}\email{}

\author[1]{\fnm{Carlo A.} \sur{Gaggioli}}\email{}

\author[1]{\fnm{David} \sur{Zsolt Manrique}}\email{}

\author[2]{\fnm{Joe} \sur{Gibbs}}\email{}

\author[2]{\fnm{Sean} \sur{Harding}}\email{}

\author[1]{\fnm{Thomas M.} \sur{Soini}}\email{}

\author[2]{\fnm{Neil} \sur{Gaspar}}\email{}

\author[2]{\fnm{Robert} \sur{Harker}}\email{}

\author[2]{\fnm{Mark} \sur{Storr}}\email{}

\author[1]{\fnm{David} \sur{Mu\~noz Ramo}}\email{}

\affil[1]{\orgname{Quantinuum}, \orgaddress{\street{Terrington House, 13-15 Hills Road}, \city{Cambridge} \postcode{CB2 1NL}, \country{United Kingdom}}}

\affil[2]{\orgname{AWE}, \orgaddress{\city{Aldermaston, Reading} \postcode{RG7 4PR},  \country{United Kingdom}}}

\abstract{
Due to the wide range of technical applications of actinide elements, a thorough understanding of their electronic structure could complement technological improvements in many different areas.
Quantum computing could greatly aid in this understanding, as it can potentially provide exponential speedups over classical approaches, thereby offering insights into the complex electronic structure of actinide compounds. 
As a first foray into quantum computational chemistry of actinides, this paper compares the method of quantum computed moments (QCM) as a noisy intermediate-scale quantum algorithm with a single-ancilla version of quantum phase estimation (QPE), a quantum algorithm expected to run on fault-tolerant quantum computers. We employ these algorithms to study the reaction energetics of plutonium oxides and hydrides. In order to enable quantum hardware experiments, we use several techniques to reduce resource requirements: screening individual Hamiltonian Pauli terms to reduce the measurement requirements of QCM and variational compilation to reduce the depth of QPE circuits. Finally, we derive electronic structure descriptions from a series of representative chemical models and compute the
% reaction 
energetics from quantum experiments on Quantinuum's H-series ion trap devices using up to 19 qubits.
We find our experiments to be in excellent agreement
 with results from classical electronic structure calculations and state vector simulations.}


\maketitle
\input{introduction}

\input{methods}
\input{results_overview_new}

\input{conclusions}

\bmhead{Author contributions}
KS, CDP and DMR designed the project and conceptualised the problem. 
KS, CDP, JG, SH, TMS, NG,
RH and MS performed literature review to choose the chemical reactions and atomistic models most suitable for this study.  
KS performed classical quantum chemistry calculations and hybrid simulations on quantum emulator and hardware.
CDP built the PBC models, performed DFT and quantum chemistry simulations and some preliminary experiments on emulator.
GGD conceptualised and developed the quantum-classical workflow for the QCM4 method, in addition to preparing and performing the experiments on emulator and hardware.
CAG conceptualised the right strategy to perform classical quantum chemistry calculations and prepared and performed classical calculations and  QPE quantum experiments on emulator and hardware.
DZM conceptualised and developed the quantum-classical workflow for the QPE method, in addition to performing preliminary tests.
KS, CDP, GGD, CAG, DZM, TMS and DMR wrote the manuscript with input from all the authors.
JG, SH, TMS, NG, RH, MS and DMR provided critical feedback and supervised the research study in all its phases. 
All authors have read and approved the manuscript.

\bmhead{Conflicts of interest}
There are no conflicts to declare.

\bmhead{Data availability}
The data that support the findings of this study will be made available from the corresponding author(s) upon reasonable request

\bmhead{Acknowledgments}
The authors thank Georgia Prokopiou and Duncan Gowland for their feedback on the manuscript. The authors also thank Joshua Savory, Vanya Eccles and Isobel Hooper for assistance in the software and hardware experiments.

This study was supported by Innovate UK, part of the UK Research and Innovation (UKRI) agency [SBRI: Quantum Catalyst Fund, project no. 10107055]. The authors performed this work partially using the mat3ra platform, a web-based computational ecosystem for the development of new materials and chemicals~\cite{Exabyte}. 

%\bibliography{references}

\end{document}

\title[SI: Quantum simulation of actinide chemistry with quantum computers]{\,\,\,\,\,\,\,\,\,\,\,Supplementary Information:\\ Quantum simulation of actinide chemistry: towards scalable algorithms on trapped ion quantum computers}

\author*[1]{\fnm{Kesha} \sur{Sorathia}}\email{kesha.sorathia@quantinuum.com}
\author[1] {\fnm{Cono} \sur{Di Paola}}\email{}

\author[1]{\fnm{Gabriel} \sur{Greene-Diniz}}\email{}

\author[1]{\fnm{Carlo A.} \sur{Gaggioli}}\email{}

\author[1]{\fnm{David} \sur{Zsolt Manrique}}\email{}

\author[2]{\fnm{Joe} \sur{Gibbs}}\email{}

\author[2]{\fnm{Sean} \sur{Harding}}\email{}

\author[1]{\fnm{Thomas M.} \sur{Soini}}\email{}

\author[2]{\fnm{Neil} \sur{Gaspar}}\email{}

\author[2]{\fnm{Robert} \sur{Harker}}\email{}

\author[2]{\fnm{Mark} \sur{Storr}}\email{}

\author[1]{\fnm{David} \sur{Mu\~noz Ramo}}\email{}

\affil[1]{\orgname{Quantinuum}, \orgaddress{\street{Terrington House, 13-15 Hills Road}, \city{Cambridge} \postcode{CB2 1NL}, \country{United Kingdom}}}

\affil[2]{\orgname{AWE}, \orgaddress{\city{Aldermaston, Reading} \postcode{RG7 4PR},  \country{United Kingdom}}}

\maketitle

\input{SI}

%\bibliography{references}

%% file: introduction.tex
\section{Introduction}
Actinides play a vital role in a wide range of applications, including energy generation, medical diagnostics and imaging technologies, nuclear safety, and catalysis~\cite{peterson2024, Haschke2000}.
Despite their significance, experimental studies of actinide compounds are often limited by their toxicity and radioactivity.
Moreover, actinide-containing materials are typically characterised by strong electron correlation due to the near-degeneracy of the 5f, 6d, and 7s orbitals and significant relativistic effects due to spin-orbit coupling.
These features often render conventional computational methods, such as density functional approximations and coupled cluster theory, unreliable for accurately describing their electronic structure.
Multireference approaches and complete active space (CAS) methods provide more accurate descriptions in such cases~\cite{Kovacs2017}. 
However, their applicability is limited to moderately small systems due to their exponentially scaling computational costs~\cite{Vogiatzis2017}.

In this context, quantum computing has the potential to usher in a new era in modelling the electronic structure of actinides~\cite{McArdle2020}.
Recent advances in  quantum computing algorithms for fault-tolerant devices offer near-linear scaling in the computational cost of CAS methods~\cite{Lee2021, Lee2022}.
This enables large-scale CAS descriptions providing near-exact solutions for complex and strongly correlated chemical systems.

This study explores the application of quantum computational techniques to the investigation of the electronic structure and reactivity of actinide systems.
A particular focus of this foray into quantum computational actinide chemistry lies thereby on the chemical behavior of prototypical species like plutonium oxides and plutonium hydrides and their role in hydride-catalysed oxidation and corrosion reactions.
A detailed understanding of such processes is critical for developing efficient and safe technologies as, for example, hydride-catalysed oxidation of actinides like plutonium can lead to thermal excursions, a variety of other hazards, and ultimately a higher risk of containment failure  during storage.~\cite{haschke1998,Haschke2000, HASCHKE201744}

The quantum algorithms used in this work can be grouped into two main categories. The first category is Quantum Phase Estimation (QPE) \cite{Kitaev1995} that offers an efficient way, given a state with sufficient overlap with the true wave function, to determine the eigenvalues of a unitary by reconstructing the unknown phase to a high precision.
In quantum chemistry, QPE enables accurate computation of ground- and excited-state energies and, notably, provides a polynomial-time solution to the full configuration interaction (FCI) problem~\cite{abrams1999}. However, the practical implementation of standard QPE is hindered by its substantial circuit depth and the need for a large number of ancilla qubits; challenges that are exacerbated further by the limitations of current and near-term utility-scale quantum hardware.
To address these limitations, we employ a single-ancilla-qubit variant of QPE known as the Quantum Complex Exponential Least Squares (QCELS) technique~\cite{ding2023even}, combined with a circuit recompilation strategy, which ensures that the circuit depth remains constant with respect to the powers of the unitary operator used in phase estimation.

The other category of algorithms, quantum subspace expansion, is based on projecting the Schrödinger equation to a subspace spanned by a finite number of basis states, resulting in a low-dimensional eigenvalue problem that can be diagonalized classically~\cite{lanczos50}. In particular, a Lanczos tridiagonalized form of the Hamiltonian can be derived from the expectation values of Hamiltonian moments~\cite{hollenberg93}. Measuring these moment expectations quantum computationally and evaluating the corresponding lowest eigenvalue of the Lanczos subspace, are the key concepts of the method of Quantum Computed Moments (QCM)~\cite{vallury20}. In this work, we focus on QCM4, the expansion of the ground state energy in terms of Hamiltonian moments truncated to fourth order~\cite{hollenberg94}. This approximation describes electron correlation interactions not accessible by mean-field theories like Hartree-Fock~\cite{jones22}.

The algorithms discussed above were applied to various systems containing plutonium (Pu) using Quantinuum's H1 and H2 trapped ion quantum hardware and emulators.
Where feasible, Complete Active Space Configuration Interaction (CASCI) and CASSCF (Complete Active Space Self Consistent Field) calculations were performed to benchmark the accuracy of the quantum results. 

This paper  is structured as follows. 
We begin by describing the two quantum algorithms employed in this work: statistical QPE~\cite{blunt2023statistical} and QCM4.
Next, we outline the rationale behind our choice of model systems and, then, we present the results of our simulations, followed by a discussion of the key challenges encountered during the project.
Finally, we conclude with a summary and offer perspectives for future research.

%% file: methods.tex
\section{Methods}

In this section, we describe the two quantum algorithms employed in this work, along with key details of the quantum hardware used for their implementation.

\subsection{QPE and Hamiltonian Simulation} \label{qpe_setup}

The canonical QPE algorithm for chemical electronic structure problems utilizes the time evolution operator $U(\tau) = e^{-iH\tau}$ of the Hamiltonian $H$, where $\tau$ is time.  If $\tau$ is chosen such that the eigen-spectrum of $H\tau$ (phase angles) lies within $[0,2\pi)$ the eigenvalues of the unitary and the eigenvalues of $H$ (energies) have a one-to-one correspondence. The canonical QPE estimates the phase by applying controlled powers of the unitary time evolution operator, where the controls are on an auxiliary qubit register. The phase is then read out from this auxiliary register using an inverse Quantum Fourier Transformation (QFT). While several methods have been proposed to construct the controlled-unitary operator, such as Quantum Signal Processing~\cite{Childs2018}, the Trotter-Suzuki decomposition is the most widely used~\cite{Omalley2016,tranter2019ordering}. Trotterization, as it is commonly called, splits the exponential of a sum of non-commuting terms into a product of exponential factors. As particular example, first-order Trotterization can be expressed as $e^{-iH\tau} \approx \bigg(\prod_{j}e^{-iH_{j}\tau/r}\bigg)^{r}$, with $H=\sum_j H_j$, $r$ number of discrete Trotter steps. While exact in the infinitesimal case, $r\rightarrow \infty$, any finite number of steps introduces so-called Trotter error, which accumulates with time $\tau$ and depends on the number of terms $H_j$ and their magnitude. While larger values of $r$ reduce Trotter errors, they quickly lead to quantum circuits of prohibitively large depths for current and near-term quantum computers. This is exacerbated even more in the Trotterization of the controlled powers of unitaries needed to obtain the quantum phase with higher precision.

Due to these considerations, we focus on statistical phase estimation based on Hamiltonian simulation. This variant of QPE uses only a single ancilla qubit, reducing the canonical QPE circuit to a Hadamard-test circuit [Fig.~\ref{fig:methods} (a1)]. Canonical QPE stores a superposition of many time-evolved states with evolution times of multiples of $\tau$, and recovers the phase with a QFT, optimally with a single measurement of the ancilla register. In contrast, the statistical QPE we consider replaces the entire QFT stage with classical statistical analysis and requires multiple measurements. At each time step $t$ we independently prepare the time-evolved state and measure the complex overlap $\bra{\psi}\ket{\psi(t)}$ between the initial state $\ket{\psi}$ and its evolved counterpart $\ket{\psi(t)} = e^{-itH}\ket{\psi}$. Because only one ancilla is required, the circuit contains a single controlled time-evolution operator, dramatically decreasing the depth relative to canonical QPE.

The overall accuracy of this approach relies heavily on the effectiveness of the classical statistical analysis which is an active area of research~\cite{ding2023even, ding2023simultaneous}. In our experiments the phase is retrieved by means of the QCELS~\cite{ding2023even} method, which performs a non-linear least-squares fit to the overlaps recorded at multiple times (see Sec. S1 in the Supplementary Information (SI) for details).  

\subsubsection{Variational compilation}
While statistical QPE requires only a single application of the controlled time-evolution unitary, even a single Trotter step becomes very deep for large systems. Increasing circuit depth not only extends the simulation time but also leads to noise accumulation. Tab.~\ref{tab:2q_count} shows the typical two-qubit depth of the single step trotterized time evolution operator for the molecules we investigate. Because a single Trotter step is usually insufficient for chemical accuracy, the actual depth of a controlled time-evolution circuit can be orders of magnitude larger than the numbers in Tab.~\ref{tab:2q_count}. A practical way to bypass this prohibitive depth is variational compilation~\cite{blunt2023statistical,Tazhigulov2022, Benedetti2021,Mc_Keever_2023,Kikuchi_2023,Sun_varrecompile}. Variational compilation enables us to prepare shallow measurement circuits and thus demonstrate the potential performance of statistical phase estimation. The main goal of this prototyping effort is to evaluate how well such algorithmic combinations work for simple Pu models.

\begin{figure*}[!h]
    \centering
    \includegraphics[width=0.9\textwidth]{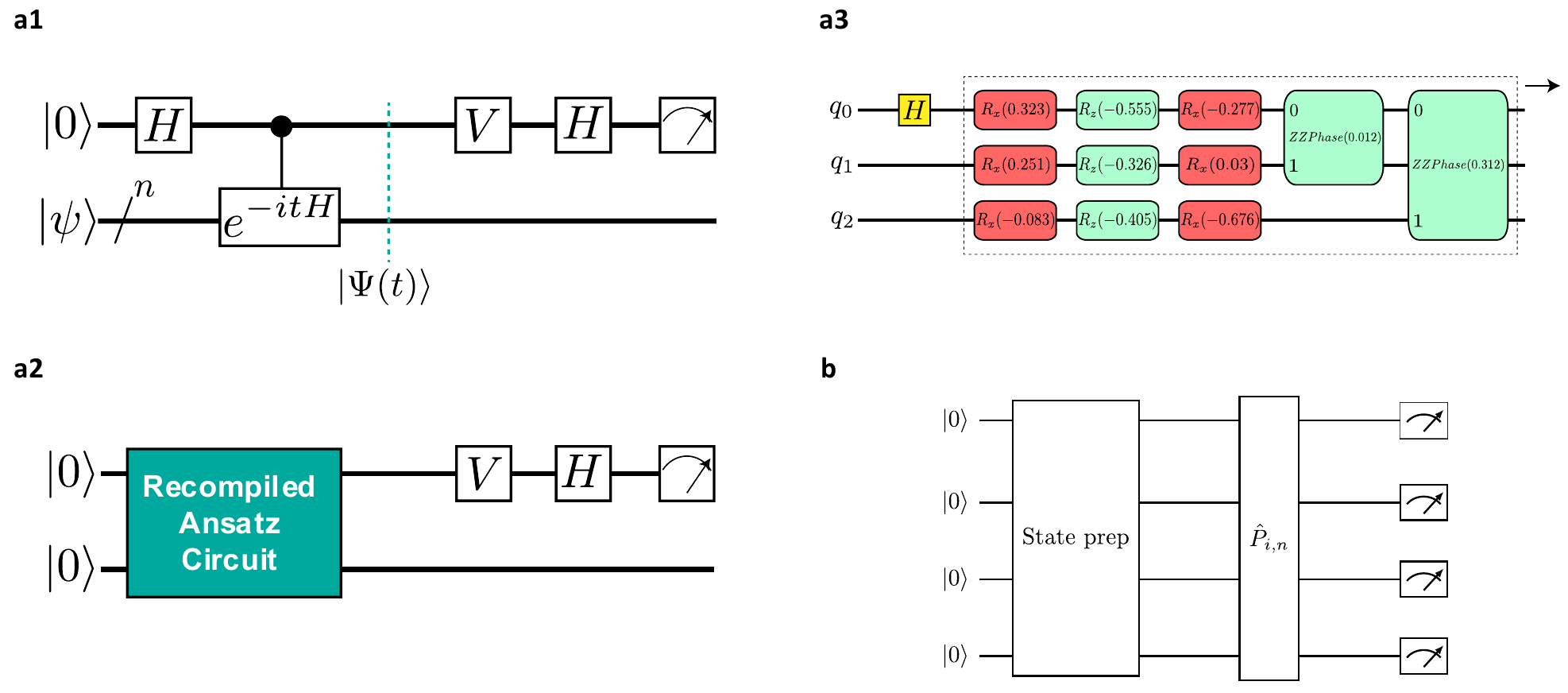}
    \caption{Circuits diagrams for QPE/QCELS and QCM4. (a1) Hadamard test that measures the complex overlap between the initial and time-evolved states, $\bra{\psi}e^{-itH}\ket{\psi}$. For the real part we set $V=I$, whereas for the imaginary part we use $V=S^{\dagger}$. The green dashed line labeled $\ket{\Psi(t)}$ marks the exact state after the controlled time evolution. (a2) Circuit schematic in which the deep controlled time evolution is replaced by a variationally compiled ansatz that prepares an approximation $\ket{\Psi}\approx\vert{\tilde{\Psi}(\theta)\rangle}$. (a3) First layer of the recompiled state-preparation circuit for the $2+1$-qubit case. Single qubit gates: Hadamard in yellow and rotation gates around the X-axis (R$_x$) and Z-axis (R$_z$) of the Bloch sphere in red and light green, respectively. 2-qubit gates: ZZPhase gate between ancilla qubit (q$_0$) and the qubits in the state register (q$_1$ or q$_2$). (b): Quantum circuit for the calculation of the expected value of Pauli string $i$ contributing to the Hamiltonian moment $n$ on a 4-qubit model. The qubit register is initialized to zero. Then, a state preparation circuit is applied to prepare the QCM4 input state. After preparing the input state, a Pauli string $\hat{P}_i^n$ of $\hat{H}^n$ is measured in the computational basis. This procedure is repeated for all Pauli strings (or for all sets of commuting Pauli strings, when commuting sets are used) contributing to Hamiltonian moments up to $n=4$.
    }
    \label{fig:methods}
\end{figure*}

The workings of the variational compilation method are illustrated in Fig.~\ref{fig:methods}, panels (a2) and (a3), and additional details are provided in the SI, Sec. S2. Panel (a2) shows that the ansatz replaces the deep controlled-time-evolution part of (a1), as well as the initial state preparation. Tab.~\ref{tab:2q_count} lists the number of two-qubit gates used in the ansatz for each active space, and the mean fidelities quantify how successful the compilations are. Here, fidelity is defined as $ \mid \bra{\psi} \ket{\psi(t)} \mid^2 $, and its mean is obtained by averaging over $t$. The variational compilation is limited both by the expressibility of the ansatz and by the global optimization procedure. 

In general, the computational complexity of variational compilation grows with the dimensionality of the unitary being compiled, the resulting number of parameters, and the difficulty of the associated minimization problem. Consequently, the compilation technique used here becomes impractical for large active spaces. The insights gained from these tests should motivate the development of scalable recompilation techniques for future QPE experiments.

\begin{sidewaystable}
    \centering
    \small
   \begin{tabular}{llllllll}
\hline
    Structure & AS &$N_{qubits}$& $1$-trotter $2q$-depth & ansatz $2q$-depth& mean fidelity (\%)&min-max fidelity (\%) & $N_{parameters}$\\
    \hline
    Pu\textsubscript{2}O\textsubscript{3} & (2,2)& 3& 15 & 12 & 99.999986&99.99991-99.9999987& 75\\
    &(2,3)&5 & 410 & 24 & 99.880&99.5-99.9968&129\\
    &(2,4)&7 & 1823  & 36 &99.87&99.5-99.9996& 183\\
    &(2,5)&9 & 6013 & 48 & 99.1&97.5-99.7&237\\
    &(2,6)&11 & 14585 & 60 & 96.2&85.4-99.8982&291\\
    &(2,8)&15 & 45787 & 84 & 84.7&49.7-99.9973&399\\
    PuH\textsubscript{2}&(6,8)&9& 5847 &48 &96.3 &59.7-99.99994&237\\
    PuH\textsubscript{3}& (5,7)&8& 3149 & 42&99.98 &99.6-99.999991&210\\
    Reactant (C1) & (10,10) &19& 127622 & 108 & 6.9&1.4-19.9&507\\
    Product  (C2) & (10,10) &19& 128826 & 108 & 16.5&4.8-48.1&507\\
    \hline
    \end{tabular}
    \vspace{7pt}
    \caption{Two-qubit depth of the controlled time evolution operator with a single trotterization step for various active space sizes. The ansatz is a hardware efficient circuit with a number of repetitions of the basic unit equal to 6. Ansatz $2q$-depth is the depth calculated in terms of two-qubit gates for the ansatz used in the variational compilation of systems in Fig \ref{fig:methods} in their active spaces AS and the mean fidelities indicate  how accurately the compiled states approximate the time evolved states across the whole time range. The numbers of variational parameters for the ansatz are also reported.}
    \label{tab:2q_count}
\end{sidewaystable}

\subsection{Subspace methods: Quantum Computed Moments} \label{qcm_method}

QCM4, the fourth order expansion of the quantum computed moments method~\cite{hollenberg94, jones22}, is used as an example of quantum algorithms suitable for near term intermediate scale quantum computers.

Like variational quantum algorithms, QCM4 relies on the same type of measurements of expectation values of Pauli strings.
However, QCM4 is a quantum subspace expansion algorithm. As such it does not rely on variational optimization of a parameterized ansatz to obtain an accurate approximation to the ground state energy, which can be plagued by barren plateau effects.
Compared to QPE, QCM4 does not require the application of a time evolution operator and therefore results in significantly shallower quantum circuits.

However, the drawback of QCM4 is its generally large number of measurement samples
(also called ``shots'') required to obtain accurate results, especially for large molecules and states.
This is due to the need to measure the expectation values of different moments of the Hamiltonian operator, i.e. $\langle\hat{H}\rangle$ as well as $\langle\hat{H}^{2}\rangle$, $\langle\hat{H}^{3}\rangle$ and $\langle\hat{H}^{4}\rangle$.
Investigating the differences between QCM4 and QPE and their effect on the quantum computed energies for Pu molecules of different sizes is the main aspect of the algorithmic comparisons in this paper. 

In the QCM4 method, expectation values of powers of the Hamiltonian operator $\langle{\hat{H}^n}\rangle$ are measured with respect to a simple initial guess for the molecular wavefunction, which we refer to as the QCM4 input state $\ket{\Psi_{\text{QCM4}}}$.
An approximation to the ground state energy is then obtained~\cite{jones22} in terms of these expectation values, using Eq.~\ref{qcm4_equation}

\begin{equation}\label{qcm4_equation}
    E_{\text{QCM4}} = c_1 - c_2\frac{c_2^2}{c_2^3 - c_2c_4} \left[ \sqrt{3c_3^2 - 2c_2c_4} - c_3 \right],
\end{equation}

\noindent where the cumulant functions $c_n$ are related to these $\langle{\hat{H}^n}\rangle$ values via

\begin{equation}\label{qcm4_equation_2}
    c_n = \langle{\hat{H}^n}\rangle - \sum_{p = 0}^{n - 2} \binom{n - 1}{p} c_{p + 1} \langle{\hat{H}^{n - p - 1}}\rangle.
\end{equation}

\noindent For Pu$_2$O$_3$ the input state $\ket{\Psi_{\text{QCM4}}}$ is a 2-electron, 2-determinant wavefunction consisting of the Hartree-Fock determinant plus its closed shell double excitation, which can be prepared using 3 2-qubit gates~\cite{greenediniz25multiconfig}.
The procedure for running QCM4 is represented in Fig.~\ref{fig:methods}, panel (b); one such circuit can be measured for each Pauli string contributing to $\hat{H}^n$ for $n=1, 2, 3, 4$.
By grouping Pauli strings into commuting sets, the number of circuits to measure for moment $n$ is significantly less than the number of Pauli strings $L_n$.
For more technical details on the use of commuting sets to measure Hamiltonian moments in relation to QCM4, we refer the reader to works on this method~\cite{jones22, vallury20}.

To further reduce the quantum resource requirements, a simple truncation procedure is applied to remove Pauli strings with coefficients below a threshold value in each $\hat{H}^n$.
This was carried out in the H1-1 experiment for Pu$_2$O$_3$ (see Sec. \ref{qcm4_2}). 

Due to the shallow input state used for QCM4, the main source of error is likely due to the noise from measuring individual contributions of Pauli strings $\hat{P}_{i,n}$.
The latter contribute to the Hamiltonian moments as

\begin{equation}
    \hat{H}^n = \sum_i^{L_n} a_{i,n}\hat{P}_{i,n}\,.
\end{equation}

\noindent To investigate this, and to possibly determine the Pauli strings whose measurement contributes the most to noise, the ideal values of each $\langle{\hat{P}_{i,n}}\rangle$ with respect to $\ket{\Psi_{\text{QCM4}}}$ are determined classically, to compare the noisy measured result to its ideal value.
We observe that for PuH$_2$ and PuH$_3$ only those Pauli strings consisting of $Z$ rotations (with no $X$ or $Y$) were non-zero, likely due to the high spin multiplicity combined with the state-average CASSCF procedure to obtain the appropriate spin orbitals within the small active space (discussed in Sec. \ref{sec:molecularhydrides}).
For Pu$_2$O$_3$, in addition to the $Z$ strings the following 4 are also non-zero: ($Y0$ $X1$ $X2$ $Y3$), ($X1$ $X2$ $Y3$ $Y4$), ($Y1$ $Y2$ $X3$ $X4$), and ($X1$ $Y2$ $Y3$ $X4$), where $X 0$ denotes an $X$ gate applied to qubit 0, \textit{etc}.
Hence, Pauli strings that provide no numerical contribution to expectation values with respect to $\ket{\Psi_{\text{QCM4}}}$ can be discarded for the purposes of calculating the energy.
This leads to vast reductions in the number of required measurements, reported in Sec. \ref{sec:qcm4_scaling}.
We refer to moments with Pauli strings removed by this procedure as being ``$\hat{P}_{i,n}$ filtered''.

\subsection{Quantum hardware, software and emulators}

We have performed our experiments on the H1-1 and H2-1 Quantinuum quantum computers. These computers have a quantum charge-coupled device (QCCD) ion trap architecture, in which the qubits are encoded as hyperfine states of electrostatically trapped Ytterbium (3+) ions~\cite{pino2021trappedionh1, moses2023racetrackh2}.
The quantum gates acting on these qubits are implemented via a series of microwave laser pulses located on specific points of the chip, the so-called gate zones.
Whenever a gate operation is applied to one or two qubits, the corresponding ions are shuttled to a scheduled gate zone and exposed to the microwave pulse corresponding to the gate.
The main features of this hardware are:

\begin{itemize}
    \item All to all connectivity between qubits, which prevents the use of extra SWAP gates (and extra noise) to implement complex circuits.
    \item Two-qubit fidelities above 99.8\%.
    \item Mid-circuit measurements and qubit reuse.
    \item Long coherence times for the qubits.
\end{itemize}

H1-1 and H2-1 Quantinuum devices provide 20 and 56 qubits, respectively. There are also classical emulators available, called H1-1E and H2-1E, which can recreate, to high fidelity, noisy experiments of the devices. 

All quantum computational calculations were performed using a development version of our in-house software suite, InQuanto~\cite{inquanto}.

%% file: results_overview_new.tex
\section{Results and Discussion}

In this section we present and discuss the results obtained from QPE simulations and QCM4 measurements with emulator (H1-1E) runs as well as quantum experiments on the H1-1 quantum computing system.
But first, we introduce the atomistic models used in the present study.

\subsection{Classical models and reference calculations}\label{sec:system}

\begin{figure*}[!t]
    \centering
    \includegraphics[width=0.8\textwidth]{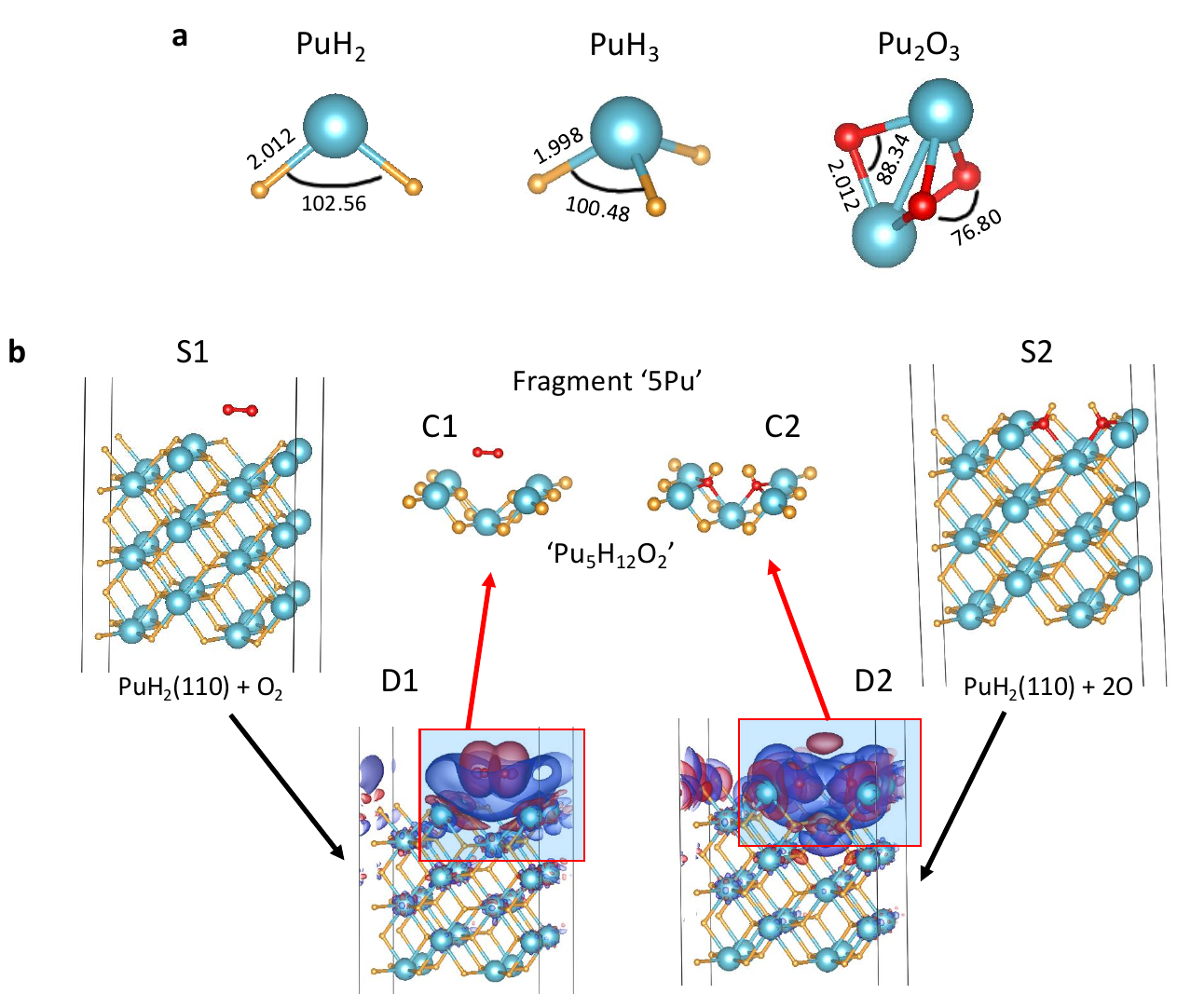}
    \caption{Plutonium, hydrogen and oxygen are represented by light blue, yellow and red spheres, respectively. (a): Structural arrangements (bond lengths in \AA, angles and dihedrals in degrees relaxed at the DFT level) of plutonium dihydride (PuH$_2$), trihydride (PuH$_3$) and plutonium sesquioxide (Pu$_2$O$_3$) isolated fragments. (b): Structural arrangements of: \textbf{S1}, \textbf{S2}) O$_2$ and atomic oxygen on PuH$_2$ (110) fluorite-like face-centred cubic (FCC) surface, respectively, in periodic boundary conditions (PBS). \textbf{D1}, \textbf{D2}) periodic supercell depicted with the difference in total electron density (DFT) as $\Delta \rho = \rho(PuH_2 + O_2/ 2O) - \rho(PuH_2) - \rho(O_2 / 2O)) $ where $\rho$ is the total density. Red and blue volumetric data with isovalue 10$^{-4}$ [$e$/bohr$^{3}$] represent gain and loss in $\rho$, respectively. \textbf{C1}, \textbf{C2}) 5 plutonium atoms clusters (19 atoms in total, also called 5Pu or Pu$_5$H$_{12}$ model) obtained by cutting a smaller fragment (area delimited by red boxes) from the periodic surface. C1 model also referred to as Pu$_5$H$_{12}$+O$_2$ or O$_2$@PuH$_2$ and C2 as Pu$_5$H$_{12}$+2O or 2O@PuH$_2$ in the main text.
    }
    \label{fig:struct_models}
\end{figure*}

We aim to model two hypothetical plutonium reactions; Firstly, the oxidation of $\text{PuH}_2$ in the gas phase $(g)$ as an analogue for the oxidation of plutonium hydride in the solid state~\cite{BALASUB2007,zhang2018}.
\begin{equation}\label{eq:mol_reac}
    3\text{PuH}_2(g) + \frac{3}{4} \text{O}_2(g) \rightarrow \frac{1}{2} \text{Pu}_2\text{O}_3(g) + 2\text{PuH}_3(g)
\end{equation}
We generated a representative set of orbitals for the sesquioxide Pu$_2$O$_3$ and the hydrides PuH$_2$ and PuH$_3$ (structures depicted in Fig.~\ref{fig:struct_models}, panel (a)).
With these orbitals, we defined subsets of them (the so-called active spaces or AS) to balance description of the chemistry of each model with resources needed to perform the quantum calculations.
Each AS is specified by the number of electrons and the number of orbitals that contribute to the wavefunction of the chemical model considered.

The second reaction examined in this paper is the dissociation of physisorbed $\text{O}_{2}$ on a $\text{PuH}_2$ surface $(s)$~\cite{Smith2022} defined by cuts along the crystallographic (110) direction
\begin{equation}\label{eq:solid_reac}
    \text{O}_2@\text{PuH}_2(110)(s) \rightarrow 2\text{O}@\text{PuH}_2(110)(s).
\end{equation}
To this end, we first performed (at the DFT level) the geometry relaxation of a single O$_2$ molecule (\textbf{S1}, Fig.~\ref{fig:struct_models}, panel (b)) and two oxygen atoms (\textbf{S2}, Fig.~\ref{fig:struct_models}, panel (b)) on top of a periodic crystalline PuH$_2$(110) surface model. 
We then truncated the periodic structures to small non-periodic clusters with stoichiometry Pu$_5$H$_{12}$O$_2$, depicted in \textbf{C1} and \textbf{C2}, Fig.~\ref{fig:struct_models}, panel (b), respectively.
This was motivated by the observed weak perturbative effect of the oxygen adsorption on the electron density of the PuH$_2$ surface, affecting at most the surface and first subsurface layers (see \textbf{D1} and \textbf{D2}, Fig.~\ref{fig:struct_models}, panel (b)). More technical information about the creation of these models is provided in the Sec. S3 of the SI.

An analysis of various spin states carried out at the DFT level in the PuH$_2$, PuH$_3$ and Pu$_2$O$_3$ fragment models reveals that states with high multiplicity (6, 5 and 10 unpaired electrons, respectively) are energetically most stable.
    Similarly, for the Pu$_5$H$_{12}$O$_2$ model we found a multiplicity as high as 29 (28 unpaired electrons) most stable.
    These high spin states require all $f$-orbitals of the Pu atoms to be included in the active space, which would require an active space description of 14 qubits per Pu atom.
    Clearly, for the Pu$_2$O$_3$ structure and Pu$_5$H$_{12}$O$_2$ surface model this requirement exceeds our available quantum hardware resources.
    In contrast, using singlet multiplicity in our classical CASSCF simulations (please refer to Sec.~\ref{surface O2 dissociation}) shows reasonable convergence of the O$_2$-dissociation energy on the Pu$_5$H$_{12}$ model with respect to the active space, allowing for a description up to 20 qubits, AS(10e,10o).
    To overcome this issue, we decided to use singlet state results whenever the high-spin ground states cannot be feasibly computed and the true ground state results otherwise.
    This allows us to generate experiments for quantum hardware on all systems considered in this study.
In particular, we calculated the energies of the Pu$_2$O$_3$, \textbf{C1} and \textbf{C2} structures in a closed-shell singlet state.
In contrast, we computed the energies of the PuH$_2$ and PuH$_3$ fragment models in their high-spin ground state (5 and 6 unpaired electrons respectively) as the presence of a single Pu atom in each of these models allowed us to consider active spaces that could fit within the constraints of our hardware.

\subsection{Scaling of resources with active space sizes of Pu\textsubscript{2}O\textsubscript{3}}

In the following, we report the results for the variationally compiled QCELS and QCM4 algorithms, for the plutonium sesquioxide (Pu\textsubscript{2}O\textsubscript{3}) fragment in singlet spin multiplicity with different active space (AS) selections.
We analyzed the Pu\textsubscript{2}O\textsubscript{3} structure as a proxy for the formation of plutonium sesquioxide through the oxidation of plutonium dihydride (see Eq.~\ref{eq:mol_reac} in gas-phase).
To determine a suitable AS for our experiments, we studied the energy dependence of the Pu$_2$O$_3$ molecular model with respect to the number of selected orbitals.
This step was performed by means of classical CASCI calculations, since it allows for smoother energy change with AS size, owing to the fact that orbitals are not reoptimized and therefore do not change between AS sizes.

\begin{figure*}[!t]
    \centering
    \includegraphics[width=0.9\textwidth]{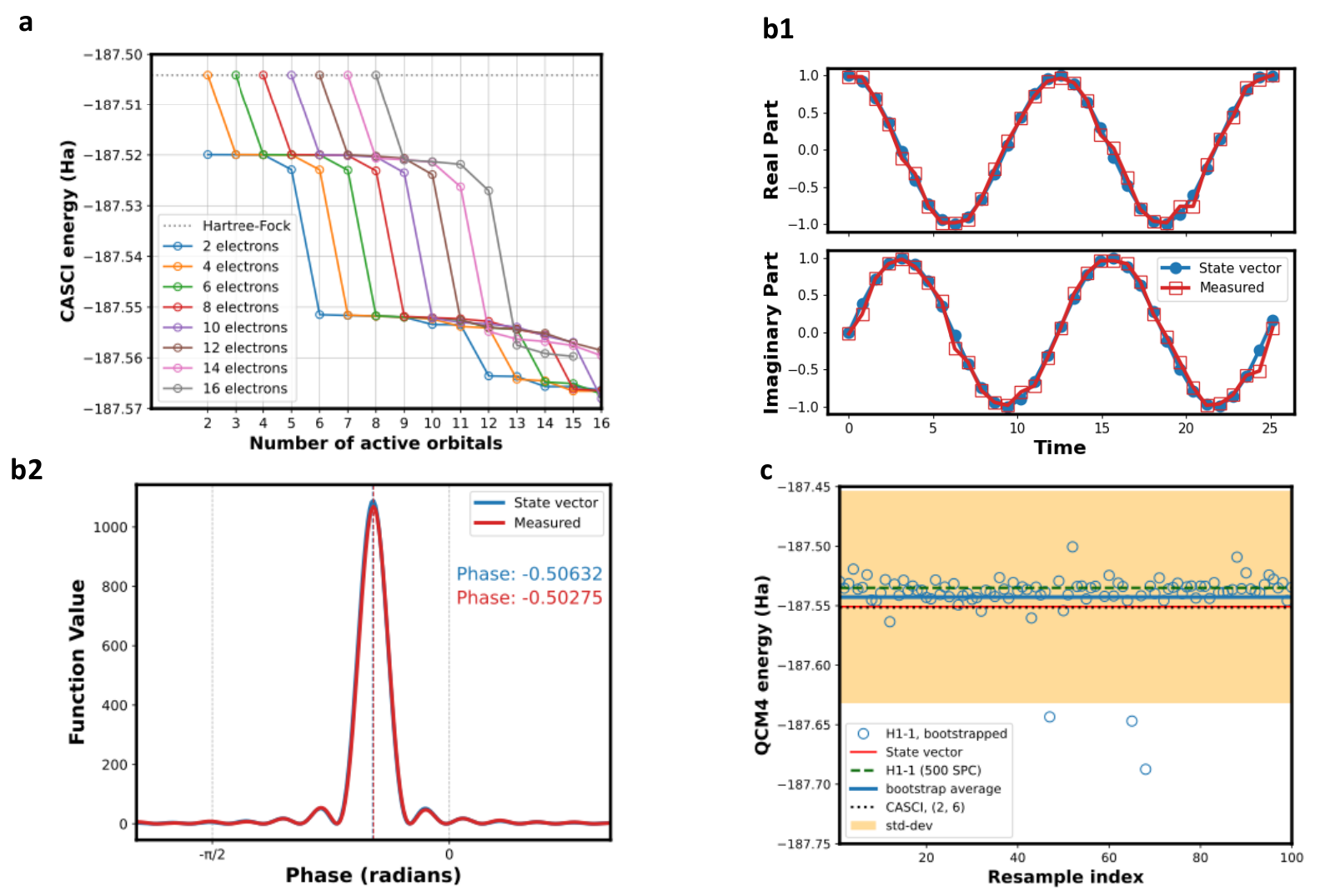}
    \caption{(a): CASCI energies of Pu$_2$O$_3$ as a function of active space size. At (2e,2o), roughly 16 mHa of correlation energy (relative to Hartree-Fock) is obtained. Large plateaus in energy are seen with increasing active space size, indicating the importance of excitations to higher virtual orbitals; (b1): Real and imaginary part of the complex overlaps $\langle\psi\vert\psi(t)\rangle$ calculated with state vector (blue curve) and measured with the hardware H1-1 with 100 SPC (red curve). The total number of qubits is 3. Values of the standard deviations over the hardware measured overlaps due to shots sampling are shown in Tabs. S1 and S2 in the SI; (b2): Phase estimation with QCELS using the overlaps from state vector (blue curve) and from measurements with the hardware H1-1 with 100 shots per circuit (red curve). The total number of qubits is 3. The x axis is the phase while the y axis represents the QCELS objective function (Eq. S6 in the SI); (c): QCM4 method applied to the active space (2e,6o), for Pu$_2$O$_3$, obtained from moments measured on the H1-1 device using 500 SPC. The measurement results are resampled using a bootstrapping technique to emulate a distribution over statistically independent device runs (100 resamples). Orange shaded region indicates the standard deviation of the resampled results.
    }
    \label{fig:pu2o3_results}
\end{figure*}

In panel (a), of Fig.~\ref{fig:pu2o3_results} we show the CASCI energy profile of Pu\textsubscript{2}O\textsubscript{3} (singlet spin state) as a function of the active space size, going from 2 electrons in 2 orbitals [(2e,2o)] until 16 electrons in 16 orbitals [(16e,16o)]. 

Focusing on the two-electron results (blue curve), we noticed that until 4 orbitals, the CASCI energy remains similar to that obtained from 2 electrons in 2 orbitals, but decreases noticeably between 5 and 6 orbitals to then reach a plateau until 9 orbitals. 
The inclusion of additional orbitals beyond an AS size of 8  therefore does not result in a substantially improved CASCI energy (see Fig.~\ref{fig:pu2o3_results}, panel (a)) but causes significantly larger computational costs (classically and quantum) for the variational recompilation. Hence, execution of the circuits beyond (2e,8o) would become very expensive compared to the gain in the absolute value of the total energy.
For these reasons, we decided to analyse the performance of the statistical QPE algorithm for active spaces up to 2 electrons in 8 orbitals. 

\subsubsection{Statistical QPE Experiments on H1-1E and H1-1}
\label{qpe_results}

In this section, we report results for the statistical QPE (with QCELS) algorithm using both emulator and hardware runs; unless mentioned otherwise, these results were obtained with 100 shots per circuit (SPC). The measurement circuits were prepared for $33$ time steps over a total simulation time that includes at least two apparent periods for the overlaps, as can be seen in Fig.~\ref{fig:pu2o3_results} panel (b1). With this choice of time step and total simulation time, the QCELS energies extracted from the state vector simulation are close to the CASCI energies as it is shown in Tab.~\ref{tab:statistical_qpe_AS}.
For each of the $33$ time steps in the QPE calculations, we variationally compiled the circuits to prepare measurement circuits for the real and imaginary parts of the overlap. For the variational compilation we used an ansatz with repeated layers of rotations and entangling gates (see Sec. S2 in the SI for additional information). Fig.~\ref{fig:methods}, panel (a3) depicts the first layer of the ansatz for a $2+1$-qubit system. For larger systems the structure is similar: each layer consists of single-qubit rotations and a cascade of 2-qubit \texttt{ZZPhase} gates (native on Quantinuum devices) that sequentially couple the ancilla to the system qubits. We found that this ansatz works reasonably well up to $11$ qubits with the allocated 2-qubit counts, yielding a mean fidelity, across all time steps, of $96.2\%$ (see Tab.~\ref{tab:2q_count}). Beyond $11$ qubits the mean fidelity drops significantly. However, the errors in the overlaps, crucial for QCELS post-processing, remain much smaller. Fig.~\ref{fig:clusters_results1} shows we performed hardware measurements up to $19$ qubits.

We initially considered an AS of 2 electrons in 2 molecular orbitals. By exploiting symmetries such as particle and spin conservation~\cite{bravyi2017}, we were able to reduce the 4 Jordan-Wigner mapped~\cite{jordan1928} qubits to 2 plus one additional ancillary qubit (three qubits in total), necessary to perform our experiments (with the total of $12$ 2-qubit gates).
We prepared our initial state by including all determinants with coefficients greater than $0.03$ in the CI expansion (after carrying out a CASCI(2e, 2o) calculation), to maximize the overlap between initial and target ground state.
For larger active spaces, we used the same $0.03$ threshold for the determinant inclusion in the initial state. In all cases the initial states have larger than $96\%$ fidelity with the exact ground state, which is sufficient for the QCELS method \cite{ding2023even}.

The overlap $\bra{\psi}\ket{\psi(t)}$ measured after 100 SPC on the quantum device is shown for the real and imaginary parts (red curve) in Fig.~\ref{fig:pu2o3_results}, panel (b1). We noticed a good agreement with the state vector (SV) results (blue curve), with only minor deviations arising from statistical noise of the hardware measurements.

The agreement with theory is also reflected in the estimated phases obtained with the QCELS approach (as shown in Fig.~\ref{fig:pu2o3_results}, panel (b2)) from the complex time series of the overlap evolution (see Eq. S9 in SI).
The phases correspond to total energies of $-187.51989$ Ha and $-187.51924$ Ha for state vector and hardware, respectively.
Our hardware experiments using 100 shots are thus very accurate.
With a deviation only $0.7$ mHa from the state vector energy, we find our quantum experiments to yield results well within the threshold of chemical accuracy.

We now proceed to analyze results for more complex active spaces. In Tab.~\ref{tab:statistical_qpe_AS} we provide an overview of the different experiments in terms of active spaces, number of qubits, 2-qubit gates depth, state vector QPE energy, energies obtained from emulator and hardware runs, and the difference between emulator (or hardware) and state vector results.
We can see that, as the active space size increases (and therefore the number of qubits and 2-qubit depths are larger), the differences in energy between state vector (SV) and emulator increase, with the maximum difference being around $6$ mHa for the active space of (2e,8o).  We attribute this effect to the ansatz that is unable to capture the full complexity of the exact state, which is also reflected by the deviation between the SV simulation with the original $\ket{\Psi}$ and variationally compiled $\vert{\tilde{\Psi}(\theta)}\rangle$  (see Figs. S1 and S2 in the SI). 

\begin{sidewaystable}
    \centering
    \small
   \begin{tabular}{lllllllll}
\hline
AS (elec,orb) & $N_{qubits}$ & $2q$-depth & E CASCI (Ha)&E SV (Ha)& E\textsubscript{emu} (Ha)& E\textsubscript{har}(Ha)& $\Delta$E\textsubscript{emu} (mHa) & $\Delta$E\textsubscript{har} (mHa)\\
    \hline
    (2,2) & 3 & 12 & -187.51993 & -187.51989  & -187.51969  & -187.51924 &0.0002&0.0007 \\
    (2,3) & 5 & 24 & -187.51995 & -187.51992  & -187.51976
    & - &0.0002&- \\
    (2,4) & 7& 36 & -187.51995 & -187.51995  & -187.52092  & - & 0.001 & - \\
    (2,5) & 9& 48& -187.52289 & -187.52289  & -187.52049  & - & 0.0024 & - \\
    (2,6) & 11& 60& -187.55148 & -187.55156  & -187.55262  & -187.55569  & 0.001 & 0.0041 \\
    (2,8) & 15& 84&-187.55178 & -187.54758  & -187.54101  & -187.54288  & 0.0065 & 0.0047 \\
    (2,8) 500 SPC & 15 & 84 & -187.55178 & -187.54758 & - & -187.54483 & - & 0.0028 \\
    \hline

    \hline
    \end{tabular}
    \vspace{7pt}
    \caption{Pu\textsubscript{2}O\textsubscript{3} (singlet) active spaces, number of qubits, number of 2-qubit gates depths, CASCI energies, energies for the state vector, emulator and hardware runs (calculations with 100 SPC when not specified), and energy differences between the emulator(emu)/hardware(har) and state vector ($\Delta$E\textsubscript{emu} and $\Delta$E\textsubscript{har}) in the QPE framework.}
    \label{tab:statistical_qpe_AS}
    
\end{sidewaystable}

For the hardware runs, we can see a similar trend as the emulator runs, namely the differences in energy between state vector and hardware increase with increasing active space, being around 5 mHa for an active space (2e,8o).
For this AS, we find that increasing the number of shots to 500 SPC reduces the difference to the state vector result back to 3 mHa, indicating the need for higher statistical averaging in this case (see also Fig. S3 in the SI).

\subsubsection{QCM4 Experiments on H1-1E and H1-1}\label{qcm4_2}

H1-1 experiments for Pu$_2$O$_3$ were performed using the largest active space accessible by this method, 2 electrons in 6 orbitals, which translates to 12 qubits.

To measure all Pauli terms required for QCM4, 2412 measurement circuits are required (after grouping the terms into commuting sets).
To reduce this quantum resource requirement, Pauli strings with coefficients below a threshold are removed after the full set of Pauli strings required for all 4 Hamiltonian moments is obtained. We find that a Pauli coefficient threshold of 0.001 provides a good approximation to the QCM4 energy, with error \(\sim \)0.1 mHa.
At this threshold, the number of Pauli circuits to be measured after grouping into fully commuting sets (as implemented in TKET~\cite{tket20}) reduces to 516. 

The corresponding results are shown in panel (c), Fig.~\ref{fig:pu2o3_results}.
To establish a standard deviation for our experiments, the hardware result is bootstrapped to emulate 100 resamples of the same measurement procedure; the resulting standard deviation amounts to 0.18 Ha (orange shaded region).
This error largely originates from two aspects related to statistical precision. First, from the large number of measureable circuits that remain even after omitting Pauli strings below the threshold and grouping into commuting sets. Second, from the relatively small number of shots (500 SPC). 

Furthermore, the $\hat{P}_{i,n}$ filtering procedure (see Sec. \ref{qcm_method}) was performed for Pu$_2$O$_3$ and this resulted in 2 circuits to measure after also accounting for commuting sets (see also Sec. \ref{sec:qcm4_scaling}).
The contributions to the QCM4 energies can be reconstructed from these 2 circuits, allowing for highly accurate QCM4 energies even in the presence of device noise and measurement sampling error (10,000 SPC was used for the Pauli filtered QCM4 results for Pu$_2$O$_3$).
This is shown by 5 independent runs of the QCM4 method on the H1-1 emulator, the results of which are reported in Tab. \ref{tab:qcm4_filtered_pu2o3}. 

As also highlighted for the hydrides in Sec. \ref{qcm4_0}, the QCM4 method generally compares well with statistical (variationally recompiled) QPE, and with the ideal classical result, once Pauli filtering is used.
The low number of measurements and relatively shallow circuits allow for a high accuracy of the emulated QCM4 experiment.
However, when all Pauli terms are measured (no filtering), measurement noise leads to larger errors in the QCM4 method, where even for smaller SPC the statistical QPE approach achieves a higher accuracy.
We also note the ability of variationally compiled QPE to reach larger active spaces, since the number of measurable Pauli terms involved in QCM4 for the (2e,8o) active space becomes too large to be tractable with current methods, as discussed further in section \ref{sec:qcm4_scaling}.

\begin{table}[h]
\centering
\small
  \caption{\ QCM4 energy of Pu$_2$O$_3$, AS (2e,6o), calculated from QCM4 by measuring circuits corresponding to filtered Pauli strings. Measurements performed on the H1-1 emulator, using 2 circuits and 10,000 SPC. Note all values are less than 0.1 mHa above the CASCI value.}
    \label{tab:qcm4_filtered_pu2o3}
  \label{tbl:example1}
  \begin{tabular*}{0.48\textwidth}{@{\extracolsep{\fill}}lll}
    \hline
    Run & $E_{\text{QCM4}}$ & Error \\
    \hline
    1 & -187.550982 & 0.000578 \\
    2 & -187.550757 & 0.000803 \\
    3 & -187.550874 & 0.000686 \\
    4 & -187.550908 & 0.000652 \\
    5 & -187.550776 & 0.000784 \\
    \hline
  \end{tabular*}
\end{table}

\begin{table*}
\small
\caption{Number of terms ($L_n$) in the QCM4 moments for $n=1,2,3,4$, for various actives spaces of Pu$_2$O$_3$. Also shown are the total number of circuits required for QCM4 when fully accounting for commuting sets of Pauli terms ($N_{circuits}$). For small active spaces, note the saturation in the number of Pauli strings for higher moments~\cite{claudino21}. For the (2,8) active space, the QCM4 moments for $n>1$ are not calculated due to large computational overhead, and hence $L_{n>1}$ are omitted.}
    \label{tab:qcm_scaling}
  \begin{tabular*}{\textwidth}
  {@{\extracolsep{\fill}}ccccccc}
    \hline
    AS (elec,orb) & $N_{qubits}$ & $L_1$ & $L_2$ & $L_3$ & $L_4$ & $N_{circuits}$ \\
    \hline
    (2,2) & 4 & 26 & 40  & 40  & 40 & 3 \\
    (2,3) & 6 & 117 & 417 & 544
    & 544 & 14 \\
    (2,4) & 8 & 360 & 3503 & 7303 & 8320 & 85 \\
    (2,5) & 10 & 875 & 21546 & 78868 & 126293 & 475 \\
    (2,6) & 12 & 1818 & 99654 & 679818 & 1650045 & 2412 \\
    (2,8) & 16 & 5793 & - & - & - & - \\
    \hline
  \end{tabular*}
\end{table*}

\subsubsection{QCM4 Scaling}\label{sec:qcm4_scaling}

The number of Pauli strings $L_1$ in the Hamiltonian $\hat{H}$ (i.e. the first moment) scales as $N_{qubits}^4$ where $N_{qubits}$ is the number of qubits. For higher moments, the commuting properties of Pauli strings contributing to $\hat{H}^n$ allow the reduction of the number of terms $L_{n}$ significantly below $N_{qubits}^{4n}$, and $L_{n}$ can even saturate as a function of $n$~\cite{vallury20, claudino21}. The precise scaling is system dependent and determined by the commutativity of individual Pauli strings comprising $\hat{H}$ and the strings resulting from product terms of $\hat{H}^{n>1}$. For the QCM4 simulations carried out on Pu$_2$O$_3$, we show in Tab.~\ref{tab:qcm_scaling} the specific numbers of Pauli terms with respect to $N_{qubits}$ and $n$, along with the corresponding number of circuits to measure when fully accounting for the commuting properties of Pauli strings. 

The large number of measurements is a major bottleneck in QCM4 calculations. Even when commutativity is taken into account, the number of circuits to measure for larger active spaces is prohibitive; for the (2e,8o) active space, $L_1$ is already 5793, which potentially translates into millions of terms for $L_{n>1}$. When considering Pauli filtering for the (2e,8o) active space, the large number of Pauli terms translates to a large number of Pauli string expectation values to evaluate classically, requiring large classical compute resources. Overall, the overhead caused by the number of Pauli terms required by QCM4 seems indeed more restrictive than the corresponding QPE simulations. The largest QCM4 calculation we report in this paper therefore corresponds to the 12 qubit, (2e,6o) active space of Pu$_2$O$_3$.

\subsection{The molecular hydrides: PuH\textsubscript{2} and PuH\textsubscript{3}}\label{sec:molecularhydrides}
Unlike plutonium sesquioxide, we found that, in order to describe the nature of the correlation of the molecular proxies of PuH\textsubscript{2} and PuH\textsubscript{3}, all plutonium f-orbitals must necessarily be included in the active space. 

A CASCI description of these systems would imply very large active spaces, beyond the reach of currently feasible quantum experiments.
For this reason, we opted for the state-average CASSCF (SA-CASSCF, with 5 roots) procedure to avoid a poor description of the virtual orbitals at the HF level and consequently to include an optimised set of f-orbitals within a smaller active space. We then used these orbitals as initial guesses for the ground states. Since the fully spin-polarized state is used, only one spin type (spin-up) is included in the active space, which results in Hamiltonian Pauli strings containing $X$ or $Y$ rotations having no effect on the energy (maintaining spin symmetry). As reported in Sec. \ref{qcm_method}, this effect on the Pauli strings continues for higher order moments of the Hamiltonian, resulting in nonzero expectation values only for those Pauli strings consisting of $Z$ rotations, for all moments required for QCM4.

\subsubsection{Statistical QPE Experiments on H1-1E and H1-1}

The hardware results for PuH\textsubscript{2} are reported in Tab.~\ref{tab:Pu2O3_formation_reaction_data}.
The active space required for a suitable description of this fragment amounts to at least 6 electrons in 8 orbitals, with a septet spin multiplicity, implying that all 6 electrons in the active space are spin-aligned.
In such a fully spin-polarized AS no opposite spin orbitals can contribute to the electronic structure description.
After removal of these orbitals, we are left with 6 electrons and 8 alpha spin-orbitals, leading to a total of 9 qubits (8 + 1 ancillary qubit needed for the phase).
The corresponding quantum computational experiment with 100 SPC yields a total energy of -71.44511 Ha, which is in nearly complete agreement with the reference energy of -71.44508 Ha from the state vector QPE simulation. Increasing the SPC count to 500 slightly increases the difference with the theoretical value, but only by 1mHa.

The same strategy was adopted for the experiments of PuH\textsubscript{3} on quantum hardware, the results of which are also shown in Tab.~\ref{tab:Pu2O3_formation_reaction_data}.
In this case the active space required to describe the electronic structure of the PuH\textsubscript{3} fragment was 5 electrons in 7 orbitals with a sextet spin multiplicity (2S+1 with S=5/2).
As before, the fully spin-polarized active space can be reduced to effectively 5 electrons and 7 alpha spin-orbitals, therefore requiring only 8 qubits (7 qubits for the wave function description + 1 ancilla qubit needed for the statistical QPE).
The state vector QPE energy is -72.00839 Ha, while the hardware run (with 100 SPC) yields -72.00522 Ha, with an energy difference of about 3 mHa between the two.
The error of this experiment therefore exceeds chemical accuracy moderately.
Increasing the shot count to 500 SPC leaves the result largely unaltered, with 4 mHa difference instead of 3 mHa with respect to the state vector result.\\

\subsubsection{QCM4 Experiments on H1-1E and H1-1}\label{qcm4_0}

\begin{figure}[!t]
   \centering
   \begin{minipage}{0.5\textwidth}
   \centering
   \includegraphics[width=0.95\textwidth]{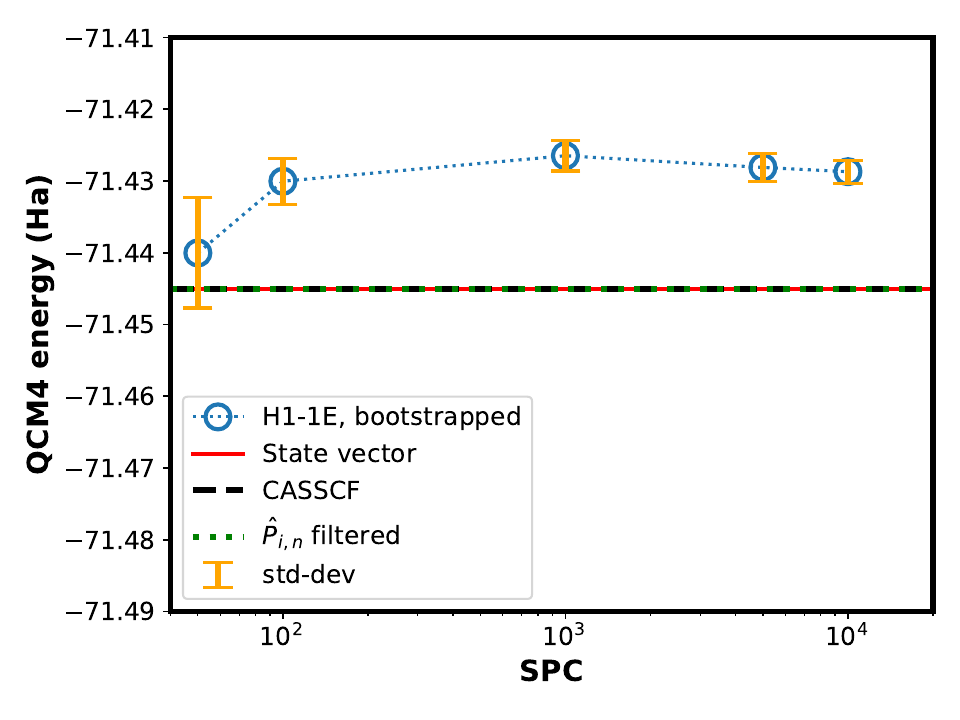}
   \end{minipage}
   \begin{minipage}{0.5\textwidth}
   \centering
   \includegraphics[width=0.95\textwidth]{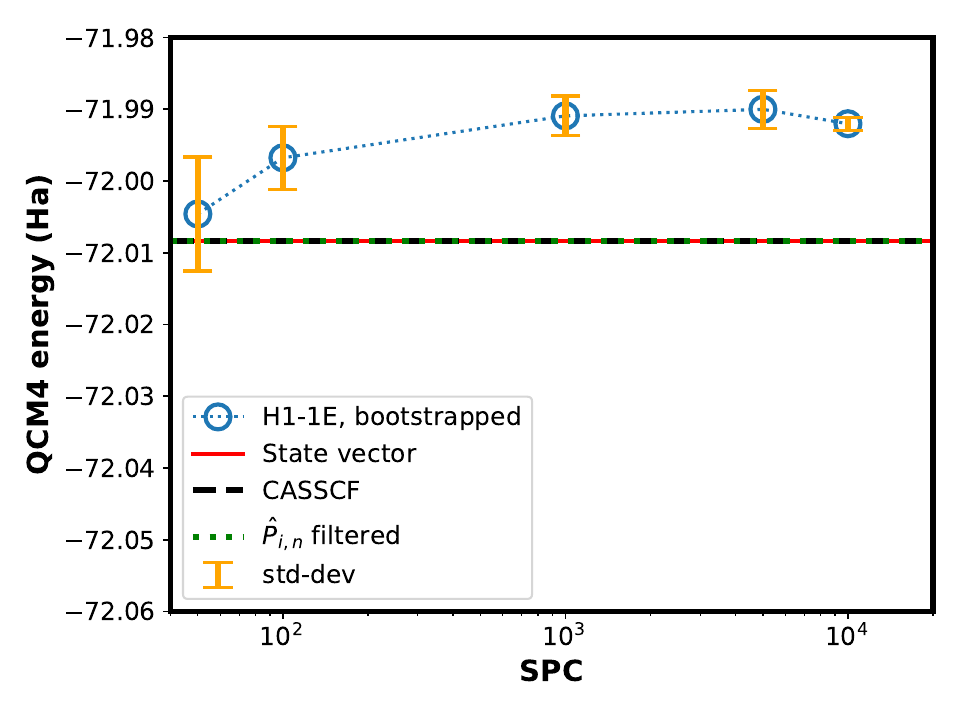}
   \end{minipage}
   \caption{QCM4 energies, with bootstrapped values obtained from the H1-1 emulator (H1-1E). 
   ``State vector'' refers to the ideal, classically evaluated value. The green dotted line corresponds to measurements of the Hamiltonian moments filtered by Pauli strings (in this case only strings of $Z$ rotations remain after filtering). Bootstrapping corresponds to 500 resamples. Blue circles correspond to QCM4 energies averaged over the bootstrapped ensemble, while orange error bars represent the standard deviations. Top: PuH$_2$; bottom: PuH$_3$.}
   \label{fig:qmc4_results}
\end{figure}

In Fig.~\ref{fig:qmc4_results} we show the results of measuring the QCM4 energies of PuH$_2$ and PuH$_3$ using the H1-1 emulator.
For these fragments, accounting for commutativity of Pauli terms (without Pauli filtering or truncating) results in 7 quantum circuits to be measured.
To investigate the distribution of errors in the QCM4 energies, the measurement results of the moment expectation values are bootstrapped to generate ensembles from the original measurement results.
For each SPC value shown in Fig.~\ref{fig:qmc4_results}, 500 resamples of the H1-1E measurements are taken. 

The results show a consistently decreasing standard deviation with an increasing number of measurement shots.
For PuH$_2$ and PuH$_3$, respectively, standard deviations of 1.6 mHa and 1 mHa are obtained at $10^4$ SPC.
The corresponding errors of these experiments amount to 16 mHa and 18 mHa, respectively. 

To obtain QCM4 energies for PuH$_3$ and PuH$_2$ from H1-1 hardware, the Pauli filtering procedure was performed before compiling the measurement circuits.
However, applying this technique resulted in all $\hat{P}_{i,n}$ containing only $Z$ strings, for both PuH$_3$ and PuH$_2$.
All these Pauli strings commute with each other, requiring the measurement of only a single circuit.
Since only the $\pm 1$ eigenvalues of the $Z$ operator are required, the result can be trivially extracted in the computational $Z$ basis (simply taking $+1$ or $-1$ according to the Z string).
Hence, the ideal energy is reliably reproduced when Pauli filtering reduces the moments to Pauli strings containing only $Z$ operators.

To summarize; applying QCM4 without error mitigation or Pauli filtering techniques to the hydrides PuH$_2$ and PuH$_3$ results in larger errors compared to statistical QPE (see values in Tab. \ref{tab:Pu2O3_formation_reaction_data}), even at $10^4$ SPC.
When Pauli filtering is used, the QCM4 yields a near-exact energy simply from the eigenvalues of the resulting $Z$ strings. The energies obtained from Pauli-filtered QCM4 therefore match the statistical QPE results very well.

\subsection{Reaction energy of O\textsubscript{2} dissociation on PuH\textsubscript{2} surface}
\label{surface O2 dissociation}

In the previous sections, we analyzed molecular-fragment systems,  as their orbital subspace is easily reduced to a size tractable with quantum computers.
However, a realistic description of the plutonium oxidation and corrosion processes requires an atomistic modeling of extended surfaces. 

In this section, we therefore study the O\textsubscript{2} dissociation on truncated and capped surface models of crystalline PuH\textsubscript{2}.
To this end, we relaxed the extended surface models of PuH\textsubscript{2} (with O\textsubscript{2} physisorbed and O\textsubscript{2} dissociated) with a periodic DFT formalism and subsequently converted them into cluster models containing 5 Pu atoms (\textbf{C1} and \textbf{C2} shown in Fig.~\ref{fig:struct_models}, panel (b)).
While the resulting electronic structures are significantly more complex than the models previously discussed, they nonetheless remain accessible to electronic structure calculation.

Based on the analysis of the electronic structure of the fragments PuH\textsubscript{2} and PuH\textsubscript{3} discussed in Sec.~\ref{sec:molecularhydrides}, we expect a very high spin multiplicity also for the electronic ground states of the cluster models of the PuH$_{2}$ surface.
As a quantum computational description of such high-spin polarizations would lead to qubit counts exceeding available   quantum resources, the total energies and corresponding dissociation energetics of these systems were obtained from a closed-shell (singlet) description of the electronic structure. Note that this is analogous to the treatment of Pu$_{2}$O$_{3}$, with the only difference that the electronic structure description is based on CASSCF-optimized orbitals rather than Hartree-Fock orbitals.

\begin{figure*}[!t]
    \centering
    \includegraphics[width=0.9\textwidth]{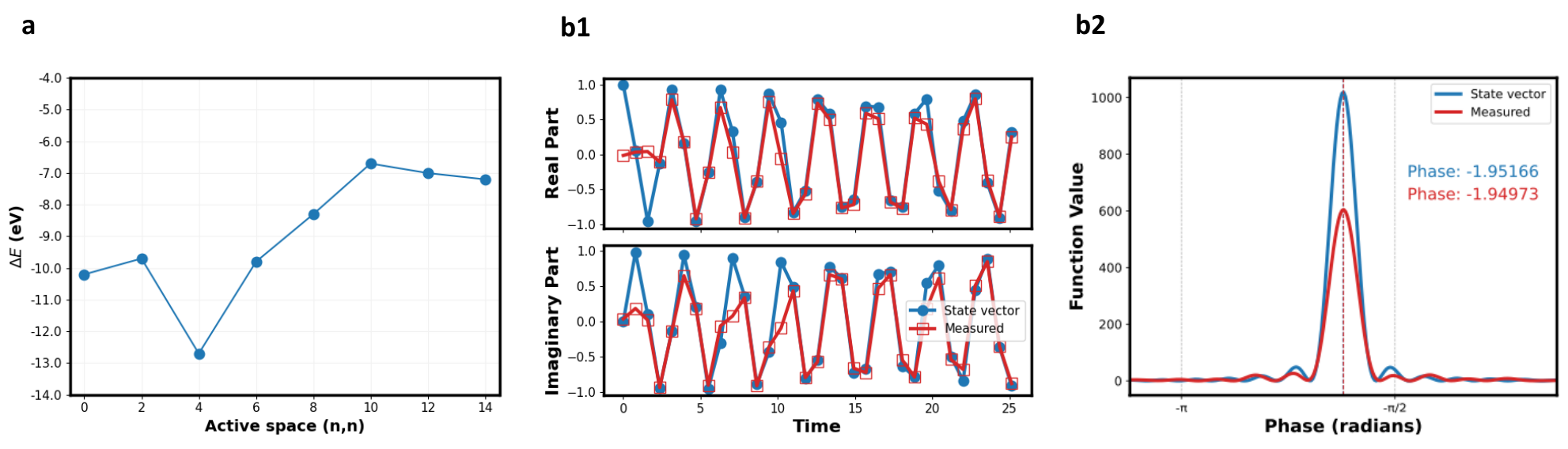}
    \caption{(a): Oxygen dissociation energy (in eV) on PuH\textsubscript{2} surfaces with respect to the active space (n electrons in n orbitals) used in classical CASSCF calculations, with n=0 corresponding to the Hartree-Fock calculation. Calculations have been carried out in the singlet spin state; (b1): cluster system 2O@PuH\textsubscript{2} real and imaginary part of the complex overlaps $\langle\psi\vert\psi(t)\rangle$ calculated with state vector (blue curve) and measured with the hardware H1-1 with 500 SPC (red curve). The total number of qubits is 19. Values of the standard deviations over the hardware measured overlaps due to shots sampling are shown in Tabs. S1 and S2 in the SI; (b2): cluster 2O@PuH\textsubscript{2} phase estimation with QCELS using the overlaps from state vector (blue curve) and from measurements with the hardware H1-1 with 500 SPC (red curve).
    }
    \label{fig:clusters_results1}
\end{figure*}

In Fig.~\ref{fig:clusters_results1}, panel (a), we show the classically computed CASSCF oxygen dissociation energies on the 5Pu clusters, i.e. $\Delta$E=E(2O$@$PuH\textsubscript{2})-E(O\textsubscript{2}$@$PuH\textsubscript{2}), obtained for different active space sizes.
We find the resulting $\Delta$E to converge to -6.7 eV at an active space of 10 electrons in 10 orbitals.

In Tab.~\ref{tab:Pu5H2_reaction_data} we report the total electronic structure energies for both reactant and product (\textbf{C1} and \textbf{C1} in Fig.~\ref{fig:struct_models}, respectively).

\begin{sidewaystable}
    \centering
    \small
\begin{tabular}{lllllllll}
  
    \hline
    Structure & \begin{tabular}[c]{@{}l@{}}AS \\(elec,orb)\end{tabular} & $N_{qubits}$ &$2q$-depth & \begin{tabular}[c]{@{}l@{}}E SV \\ (Ha)\end{tabular} &
    \begin{tabular}[c]{@{}l@{}} E\textsubscript{har} QPE \\ 100 (Ha)\end{tabular}& \begin{tabular}[c]{@{}l@{}} E\textsubscript{har} QPE \\ 500 (Ha)\end{tabular}&\begin{tabular}[c]{@{}l@{}} $\Delta$E\textsubscript{har} \\100 (Ha)\end{tabular}&\begin{tabular}[c]{@{}l@{}} $\Delta$E\textsubscript{har}  \\ 500 (Ha)\end{tabular}\\
    % Molecule & AS (elec,orb) & E CASSCF (Ha) &E QPE\textsubscript{har} (Ha)& $\Delta$E\textsubscript{har} 100 shots (mHa)\\
    \hline
    PuH\textsubscript{2} & (6,8) & 9 & 48& -71.44508 & -71.44511 & -71.44379 &0.0000 & 0.0013\\
    PuH\textsubscript{3} & (5,7) & 8 &42&-72.00839 & -72.00522 & -72.01243 &0.0032 & -0.0040\\
    \hline
    \end{tabular}
    \vspace{7pt}
      \caption{Active space used, number of qubits, state vector and QPE energies (hardware (har) runs with 100 and 500 SPC) for the hydrides PuH$_2$ and PuH$_3$. The differences in energy between QPE and SV are shown in the last two columns ($\Delta$E\textsubscript{har})}
\label{tab:Pu2O3_formation_reaction_data}
 \vspace{30pt}   
   \begin{tabular}{lllllcc}
\hline
Structure & E CASSCF (Ha) &E\textsubscript{SV} QPE (Ha) &\begin{tabular}[c]{@{}l@{}} E\textsubscript{har} QPE \\ 100 (Ha)\end{tabular} & \begin{tabular}[c]{@{}l@{}} E\textsubscript{har} QPE \\ 500 (Ha)\end{tabular} &\begin{tabular}[c]{@{}l@{}} $\Delta$E\textsubscript{har} QPE \\ 100 (mHa)\end{tabular} & \begin{tabular}[c]{@{}l@{}} $\Delta$E\textsubscript{har} QPE \\ 500 (mHa)\end{tabular} \\
    \hline
    Reactant (C1) & -389.02641  & -389.02648  & -389.02154  & -389.02560  & 4.9 & 0.9 \\
    Product (C2) & -389.27382  & -389.27408  & -389.26781  & -389.27145  & 6.3 & 2.6 \\
    \hline
    \end{tabular}
    \vspace{7pt}

    \caption{CASSCF energies, QPE energies for the state vector  (SV) and hardware (har) runs (with 100 and 500 SPC), and energy differences between the state vector and hardware ($\Delta$E\textsubscript{har}) for the reactant and product fragments of the oxygen dissociation reaction on PuH\textsubscript{2} surfaces (singlet spin state).}
\label{tab:Pu5H2_reaction_data}

\end{sidewaystable}

We compare classical CASSCF results to state vector simulations of QPE and to quantum experiments (using 100 and 500 SPC).
To this end, we also list the deviation of the quantum computed results from the state vector reference.
All these results were obtained at the active space of 10 electrons in 10 orbitals.

After taking the symmetry of these systems into account and leveraging variational recompilation, the quantum experiments comprise a total of 19 qubits (18 qubits for the state register + 1 ancillary qubit for the Hadamard test) and 2-qubit gate depths of $108$.
To the best of our knowledge, these experiments are the largest QPE calculations of a realistic electronic structure problem successfully executed on quantum hardware, surpassing previous QPE hardware experiments on one-dimensional organic chains of up to 13 qubits~\cite{Kanno2025}. However, we note that they also performed 33-qubit QPE experiments for 1-D Hubbard models.

First, we observe the state vector QPE simulations  to produce virtually identical total energies as the classical CASSCF, validating our quantum computational approach.
Comparing then state vector simulations with quantum experiments, we find a difference of around 5-6 mHa, which decreases to 0.9-2.6 mHa (see Tab.~\ref{tab:Pu5H2_reaction_data}).
This error reduction is a direct result of the higher number of samples from which the experimental result is obtained.
This increased precision at 500 SPC brings this value very close to the chemical accuracy threshold of 1.6 mHa.

Finally, we compare the reaction energies resulting from classical CASSCF calculations with those obtained from our quantum computational experiments  (with 500 SPC).
The quantum computational result of -6.69 eV deviates only by 0.04 eV (1 kcal/mol) from the CASSCF reaction energy -6.73 eV.
This demonstrates the ability of present-day QPE implementation to match the results of classical approaches in experiments executed on quantum hardware with chemical accuracy.
In Fig.~\ref{fig:clusters_results1}, panel (b1), we further report the overlaps (for both real and imaginary parts) measured after 500 SPC on the quantum device (red curve), and the overlaps calculated with state vector (blue curve) for the cluster system 2O@PuH\textsubscript{2}.
We notice a general good agreement between the two curves, with only a few outliers being noticeable.
We attribute the latter to the variational compilation procedure which becomes increasingly more approximate for higher dimensional Hilbert spaces and deeper circuits; see Fig. S5 in the SI for more detailed comparisons and  information.
As the QCELS method is a statistical approach, the individual data points do not lead to significant differences in the quantum phases obtained from simulations and quantum experiments; the total difference in phases amounts to only $0.002$, see Fig.~\ref{fig:clusters_results1}, panel (b2).

%% file: conclusions.tex
\section*{Conclusions}

The classical simulations and quantum experiments conducted in the course of this study clearly demonstrate the applicability of quantum computational chemistry to model actinide chemistry.
To this end, we successfully applied statistical quantum phase estimation and QCM4, a quantum subspace expansion algorithm, to simulate the electronic structure of actinide structures.
We employed a variational compilation method to reduce the depth of the resulting quantum circuit and ultimately enabling, to the best of our knowledge, the largest QPE experiment performed on quantum hardware for a quantum chemistry use-case so far, by leveraging a total of 19 qubits and up to 500 SPC.
A complementary picture emerges from the simulation results obtained with QCM4, where we are able to obtain accurate energies with circuits that are shallow compared to QPE.
With increasing qubit sizes, however, QCM4 leads to a steep increase in the number of measurements during the quantum experiments, due to a large number of Pauli terms.
Furthermore, a larger number of qubits and more intricate electronic structure problems will require more complex state preparation circuits.
As the circuit depth approaches the limit of coherence times of the quantum device, this could potentially render QCM4 results more sensitive to noise unless error mitigation strategies are applied~\cite{jones22}.
In general, the theoretical limits of error in quantum subspace methods is an active area of research~\cite{kirby24}, and in this study, the description of electronic structures of Pu$_2$O$_3$ with active spaces beyond (2e,6o) turned out to be a threshold prohibitively expensive to surpass.

In comparison, when leveraging variational compilation the larger (2e,8o) active space was still accessible in statistical QPE experiments.
This indicates that the measurement overhead of QCM4 is more restrictive in this case.
We also note that further research into reducing the number of required measurements could enable larger applications of QCM4 on quantum hardware. To this end, we mention previous work that shows the dependencies of these methods on the quality of their input state, which can be assessed by preparing circuits corresponding directly to linear combinations of electronic configurations~\cite{greenediniz25multiconfig}. When such input states are sparse (number of configurations $\in o(\frac{2^{N_{qubits}}}{N_{qubits}})$), the measurement overhead of QCM4 or the runtime of QPE, could be significantly reduced with only a small overhead in the circuit depth for the input state.

In more general terms, the application of these quantum chemistry methods to problems of practical interest requires the capabilities of future quantum computing systems.
The combination of hardware advancements, quantum error correction protocols and algorithmic techniques like qubitization~\cite{Low2019qubitization,Berry2018,Poulin2018}, stochastic Trotter compilation~\cite{Campbell2019,kivlichan2019randomizedhamiltonians,Wan2022} and other methods~\cite{Babbush2018,Babbush2018a} may enable quantum experiments with the deep circuits resulting from electronic structure problems of realistic actinide chemistry models.

%% file: SI.tex
\section*{S1 Energy Estimation via QCELS}
\label{sec:QCELS}
For an $n$-qubit electronic problem we write the Hamiltonian as a sum of Hamiltonian fragments 
\begin{equation}
H= h_{0}\,I + h_{1}\sum_{f} H_{f},
\end{equation}
where each $H_{f}$ is a linear combination of mutually commuting Pauli strings. The scalings  
\begin{equation}
h_{0}=\frac{1}{2^{n}}\,\text{Tr}(H),
\qquad
h_{1}=\frac{4}{\pi}\vert \vert H-h_{0} I \vert \vert
\end{equation}
ensure that the reduced operator
$\tilde H := \sum_{f}H_{f}$ has eigenvalues in $\left[-\frac{\pi}{4},+\frac{\pi}{4}\right]$. If the ground state $\vert{\psi_{0}\rangle}$ is known,
\begin{equation}
Z(t)=\bra{\psi_{0}}e^{-it\tilde H}\ket{\psi_{0}}=e^{-it\theta_{0}},
\end{equation}
and the ground-state energy is
\begin{equation}
E_{0}=h_{0}+h_{1}\theta_{0}.
\end{equation}
however typically the ground state is not known and an initial state is a $\ket{\psi}$ for which $\vert \bra{\psi_0} \ket{\psi} \vert \gg0$. Following the QCELS procedure we sample the time 
$t_{n}=n\tau$ with $n=0,\dots ,N-1$ and fixed $\tau$.  With measuring
\begin{equation}
Z_{n}=Z(t_{n})=\bra{\psi}e^{-it_{n}\tilde H}\ket{\psi},
\end{equation}
we use the QCELS objective
\begin{equation}
f(\theta)=\left\vert\sum_{n=0}^{N-1} Z_{n}\,e^{in\tau\theta}\right\vert ^{2},
\end{equation}
whose maximiser $\theta^{\ast}$ yields the energy estimate
\begin{equation}
E^{\ast}=h_{0}+h_{1}\theta^{\ast}.
\end{equation}
Each $Z_{n}$ is obtained with a single-ancilla Hadamard test, in which we prepare  
\begin{equation}
\ket{\Psi(t_{n})}=
\frac{1}{\sqrt{2}}\bigl(\ket{0}\otimes\ket{\psi}
                      +\ket{1}\otimes U(\tau)^{n}\ket{\psi}\bigr),
\end{equation} where $U(\tau)=e^{-i\tau \tilde H}$
and we measure the real and imaginary parts via
\begin{equation}
\mathfrak{R} Z_{n}=\bra{\Psi(t_{n})}X\otimes I\ket{\Psi(t_{n})},
\qquad
\mathfrak{I} Z_{n}=\bra{\Psi(t_{n})}Y\otimes I \ket{\Psi(t_{n})}.
\end{equation}
\subsection*{S1.1 Trotterization costs for the time evolution operator} 
We define the costs as the 2-qubit circuit depth. We approximate the time evolution operator with a first-order Trotter approximation,
\begin{equation}
U(\tau)=e^{-i\tau\tilde H}\approx\prod_{f}
                                   e^{-i\tau h_{1}H_{f}},
\end{equation}
where the fragments are ordered by ascending operator norm in the Trotter product, that is, the largest norm is applied to the reference first. To prepare $\ket{\Psi(t_{n})}$ we would need the controlled time evolution operator for time $t_n$, that is we would need to evaluate cost for the circuit of controlled-$U(\tau)^{n}$. Since the cost is proportional to $n$, we only calculated cost for the circuit of a single trotter step $\prod_{f}e^{-i\tau h_{1}H_{f}}$. For the costs in Tab. 1 of the main text, the circuit optimizations are performed with \texttt{pytket}, using \texttt{TermSequenceBox} and greedy optimization, \cite{paykin2023pcoast} and fully commuting sets determined by largest-first graph coloring method. \cite{tket20, inquanto}

\section*{S2 Variational compilation for hardware experiments}
To demonstrate the capability of QCELS on realistic fragments, we first carried out exact time evolution with state vector simulation: we computed each $\ket{\Psi(t_{n})}$ state vector and evaluated the corresponding $Z_{n}$ values for the QCELS objective function.  

For hardware runs we variationally compiled every exact state $\ket{\Psi(t_{n})}$ into the ansatz \begin{equation}
\ket{\Tilde{\Psi}(\vartheta)}
=U_{\mathrm{1q}}^{(L)}\left[\prod_{l=0}^{L-1} U_{\mathrm{ent}}^{(l)}\,U_{\mathrm{1q}}^{(l)}\right]\ket{+}\otimes\ket{0..0}
\end{equation} where 
\begin{equation}
U_{\mathrm{1q}}^{(l)} = \prod_{q=0}^{n_q-1} R_X^{(q)}\!\big(\vartheta^{(1)}_{q,l}\big)\,
                         R_Z^{(q)}\!\big(\vartheta^{(2)}_{q,l}\big)\,
                         R_X^{(q)}\!\big(\vartheta^{(3)}_{q,l}\big)
\end{equation}
and
\begin{equation}
U_{\mathrm{ent}}^{(l)}
=\prod_{q=1}^{n_q-1} R_{ZZ}^{(0,q)}\!\big(\vartheta^{(4)}_{q,l}\big),
\end{equation} and $n_q$ is the number of qubits including the ancilla and $L$ is the number of layers indirectly controlling the circuit depth and the number of variational parameters. An example of the beginning of ansatz circuit representation shown for the 2+1-qubit case in panel (a3), Fig. 1 in the main text.  The compilation employed \texttt{pytket}–Qujax/InQuanto together with the \texttt{optax} Adam optimiser for 500 iterations and randomized initial parameters, minimising the objective $\bigl\lVert\ket{\Psi(t_{n})}-\vert{\tilde\Psi(\boldsymbol{\vartheta})}\rangle\bigr\rVert^{2}.$
The variational compilation was performed independently for each $t_{n}$, using the same ansatz, so the circuit-resource profile is identical across all time steps, while the optimal parameters $\boldsymbol{\vartheta}$ vary with $t_{n}$.

\section*{S3 DFT geometry optimization of structures and periodic systems}
Based on previous structural studies on plutonium hydrides~\cite{BALASUB2007} and oxides,~\cite{zhang2018} for our models of PuH$_x$, with $x=2,3$, and sesquioxide Pu$_2$O$_3$, we built an ab initio model by carrying out simulations in gas phase by using the Quantum Espresso (QE) Density Functional Theory (DFT) package.~\cite{QE2009, QE2017} We performed non collinear spin-orbit coupling (SOC) calculations with DFT-D3 van der Waals dispersion correction~\cite{grimme3} for geometry optimization of structures of interest. In all DFT calculations, we employed the Perdew-Burke-Ernzerhof (PBE) GGA exchange correlation functional~\cite{Perdew1996} together with projector-augmented wave (PAW) pseudopotentials.~\cite{Blochl1994} We used the wave function and charge density cut-offs of $75$ Ry and $600$ Ry, respectively, together with a Brillouin zone sampling at the $\Gamma$ point and a Marzari-Vanderbilt smearing of $0.005$ Ry. The number of Kohn-Sham states was increased by approximately 20\% to assist with convergence of the SOC simulations. At this level of theory, we found a reaction energy of $-9.286$ eV.
 
We also built a supercell (1x2x6) for cubic Fm-3m PuH$_2$ (110) periodic surface, since it was suggested as the most stable reactive surface system by Smith et al.~\cite{Smith2022} This model was geometrically relaxed. We then generated the initial and final states by depositing O$_2$ and 2 oxygen atoms on top of the surface, respectively (structures S1 and S2, panel (b), Fig. 2 in the main text, inspired by the most stable model \textit{3P1} found in reference: Shi et al.~\cite{shi2022}). The geometries of the adsorbates and the first two layers of the surface were further relaxed. 

We adopted a DFT setup similar to that used for the simulations described at the beginning of the present section. In this particular case, we needed to correct the PBE functional by applying an Hubbard energy $U=4.88$ eV to the plutonium f-orbitals. The $U$ was derived from scratch by means of the Density Functional Perturbation Theory (DFPT) tool~\cite{timorov2022} included in the Quantum Espresso package and with a non-polarised PBE+SOC DFT calculation. We finally found a lattice parameter $a=5.380$ \AA\, in accordance with an experimental value $a=5.395 $ \AA\, (as reported by Zheng et al.~\cite{zheng2014} together with theoretical calculations).

Since the entire periodic model was extremely complex to be treated at the Hartree-Fock (HF) level of theory with the basis set of choice, we studied how the electron density of the surface was modified by the presence of oxygen atoms. In this way, we identified a smaller active region that was cut from the surface as an isolated cluster of 5 plutonium atoms (structures C1 and C2, panel (b), Fig. 2 in the main text). The obtained clusters were electronically relaxed inside the spin-polarised framework to obtained the correct multi-spin ground state 2S+1=29, with a dissociation energy at DFT level $E_{dis}=E(C2)-E(C1)=-5.547$ eV.

For the whole set of final models, we adopted a cubic box of length between 16 and 20 \AA\, and applied the Martyna-Tuckerman correction,~\cite{Martyna1999} in order to avoid the interaction among the atoms from adjacent unit cells.

\subsection*{S3.1 Classical quantum chemistry methods}
The restricted(R)- and restricted open-shell(RO)-HF mean-field calculations as well as post-HF multi reference calculations (namely CASCI, CASSCF and SA-CASSCF) have been performed with the CRENBL basis set with relativistic effective core potential (ECP) for plutonium~\cite{Ermler91} and oxygen~\cite{Pacios85} and 6-311G** basis set for hydrogen, as implemented in the PySCF quantum chemistry suite.~\cite{sun2018,sun2020} We leave more accurate treatment of relativistic effects, such as with exact 2-component or 4-component Hamiltonians, to future work. The Newton solver (Second-order SCF solver) has been used in all calculations.\\

\section*{S4 Overlap Analysis}

\begin{figure*}[!t]
    \centering
    \includegraphics[width=0.9\textwidth]{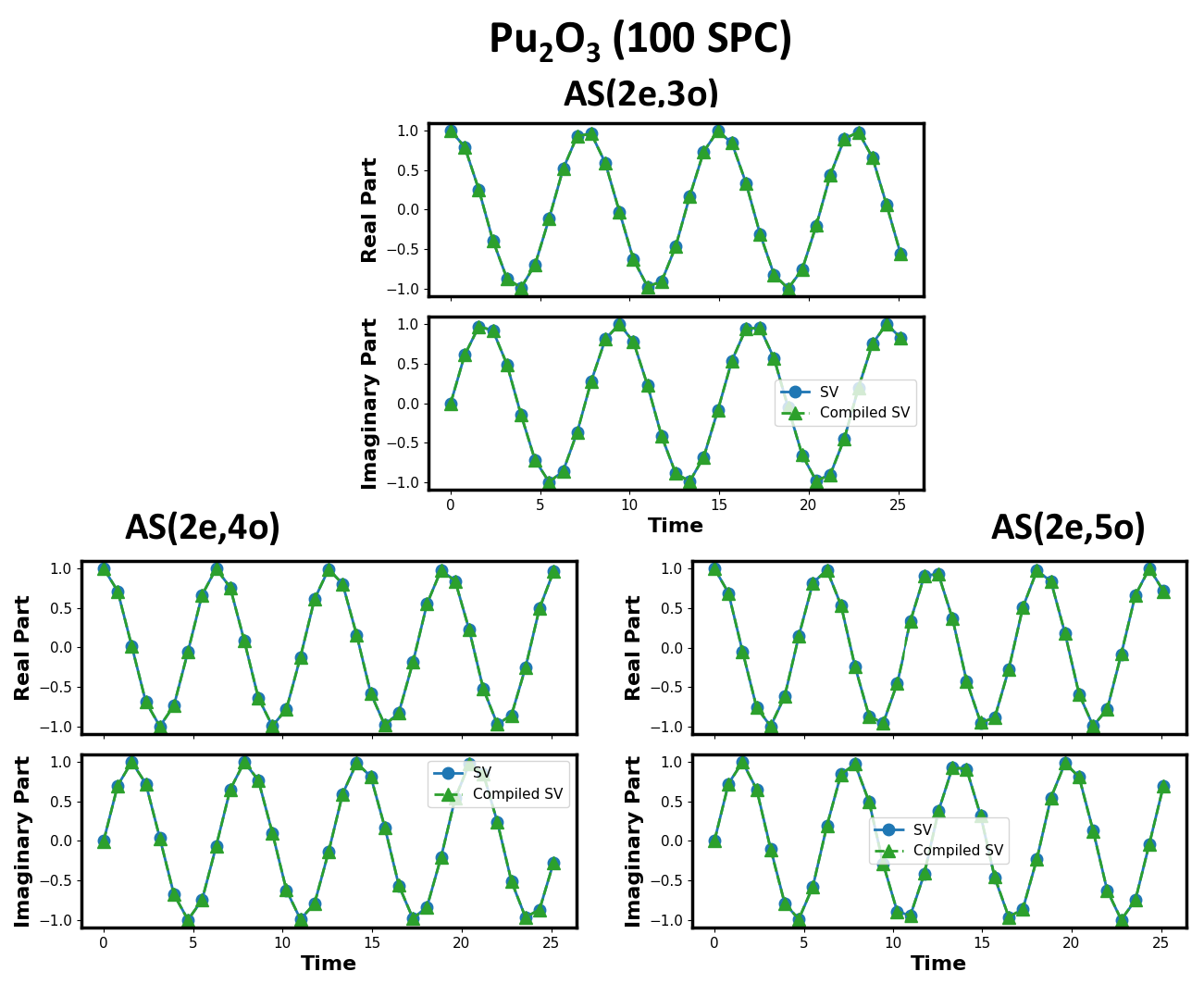}
    \caption{Pu\textsubscript{2}O\textsubscript{3} with 100 SPC: Comparison between complex overlaps $\langle\psi\vert\psi(t)\rangle$  at different time steps and active spaces  for the state vector before (SV, full blue circle) and after ansatz recompilation (Compiled SV, full green triangle). Point mismatch between state vector calculations before and after recompilation represents divergence due to approximation in the recompilation procedure.
    }
    \label{fig:pu2o3_overlaps_1_1}
\end{figure*}

\begin{figure*}[!t]
    \centering
    \includegraphics[width=0.9\textwidth]{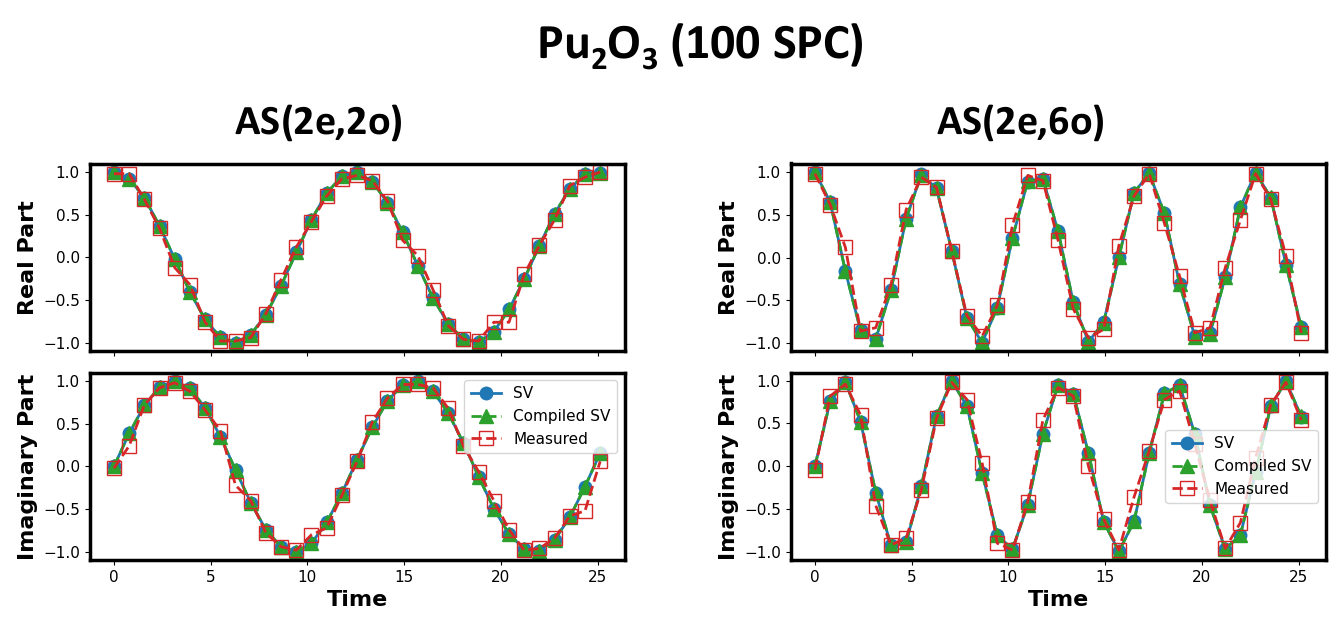}
    \caption{Pu\textsubscript{2}O\textsubscript{3} with 100 SPC: Comparison between complex overlaps $\langle\psi\vert\psi(t)\rangle$ including hardware experiments at different time steps and active spaces  for the state vector before (SV, full blue circle), after ansatz recompilation (Compiled SV, full green triangle) and measured on a hardware device (Measured, red empty squares). Point mismatch: i)  between state vector calculations before and after recompilation represents divergence due to approximation in the recompilation procedure; ii) between recompilation and hardware experiments represents divergence due to noise in the quantum measurements. Values of the standard deviations over the hardware measured overlaps due to shots sampling are shown in Tabs.~\ref{tab:stderror_rel} and~\ref{tab:stderror_img}.
    }
    \label{fig:pu2o3_overlaps_1_2}
\end{figure*}

\begin{figure*}[!t]
    \centering
    \includegraphics[width=0.9\textwidth]{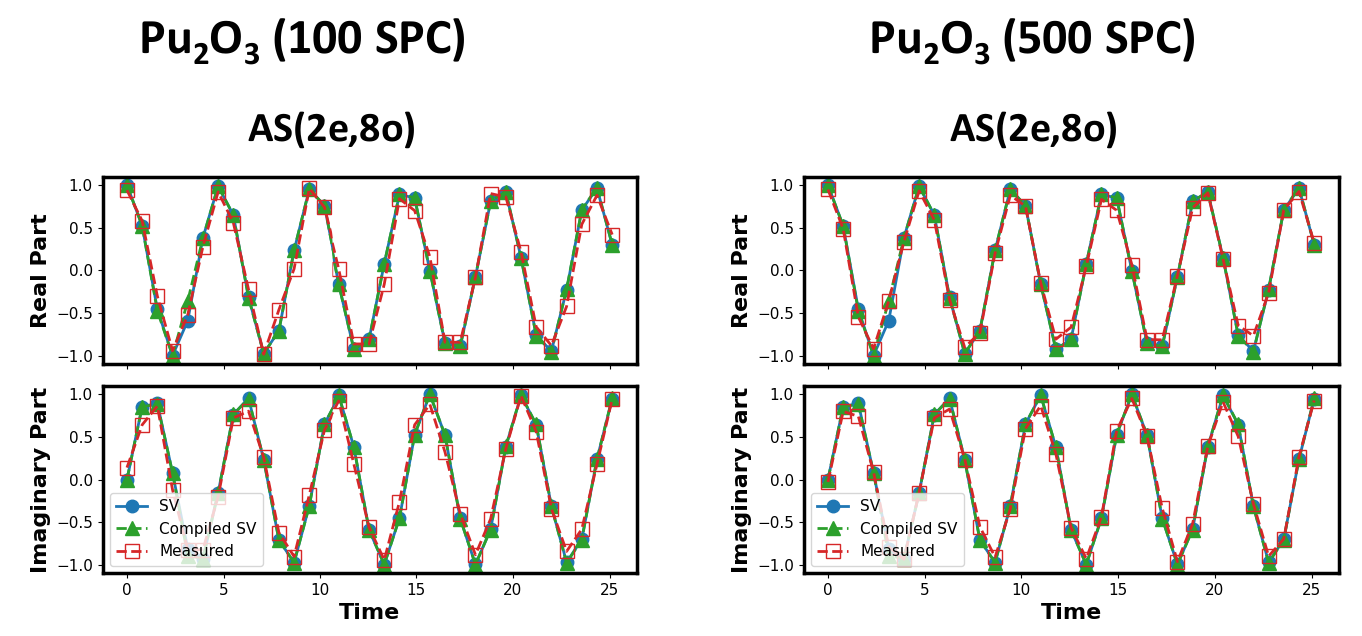}
    \caption{Pu\textsubscript{2}O\textsubscript{3} with 100 and 500 SPC for the largest complex overlaps $\langle\psi\vert\psi(t)\rangle$ including hardware experiments considered for the molecular oxide: comparison between overlaps at different time steps and active spaces  for the state vector before (SV, full blue circle), after ansatz recompilation (Compiled SV, full green triangle) and measured on a hardware device (Measured, red empty squares). Point mismatch: i)  between state vector calculations before and after recompilation represents divergence due to approximation in the recompilation procedure; ii) between recompilation and hardware experiments represents divergence due to noise in the quantum measurements. Values of the standard deviations over the hardware measured overlaps due to shots sampling are shown in Tabs.~\ref{tab:stderror_rel} and~\ref{tab:stderror_img}
    }
    \label{fig:pu2o3_overlaps_2}
\end{figure*}

\begin{figure*}[!t]
    \centering
    \includegraphics[width=0.9\textwidth]{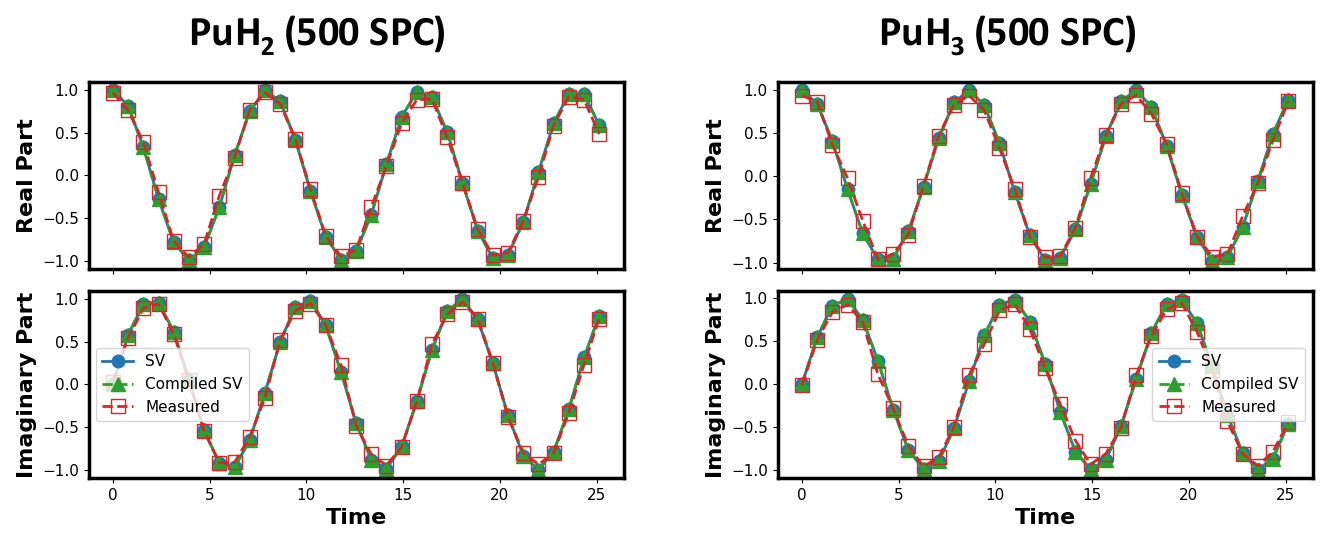}
    \caption{PuH\textsubscript{2} AS(6e,8o) and PuH\textsubscript{3} AS(5e,7o) with 500 SPC: comparison between complex overlaps $\langle\psi\vert\psi(t)\rangle$ including hardware experiments at different time steps for the state vector before (SV, full blue circle), after ansatz recompilation (Compiled SV, full green triangle) and measured on a hardware device (Measured, red empty squares). Point mismatch: i)  between state vector calculations before and after recompilation represents divergence due to approximation in the recompilation procedure; ii) between recompilation and hardware experiments represents divergence due to noise in the quantum measurements. Values of the standard deviations over the hardware measured overlaps due to shots sampling are shown in Tabs.~\ref{tab:stderror_rel} and~\ref{tab:stderror_img}
    }
    \label{fig:puh2_puh3_overlaps}
\end{figure*}

\begin{figure*}[!t]
    \centering
    \includegraphics[width=0.9\textwidth]{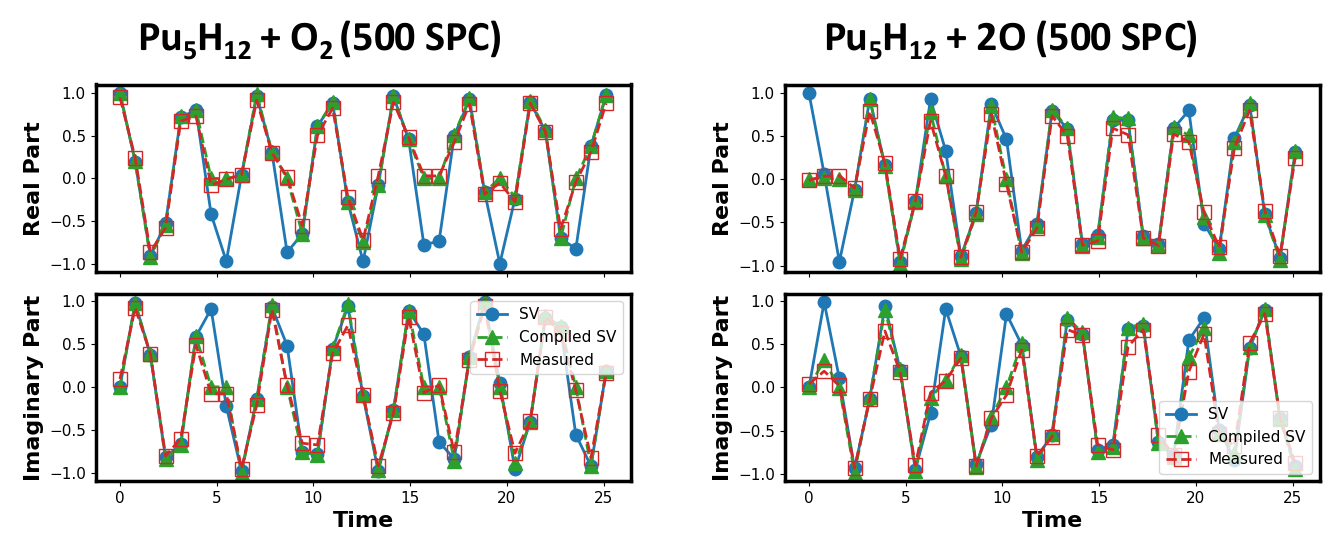}
    \caption{Cluster systems Pu\textsubscript{5}H\textsubscript{12} + O\textsubscript{2} (C1 in Fig. 2, panel (b) main text) and Pu\textsubscript{5}H\textsubscript{12} + 2O\ (C2 in Fig. 2, panel (b) main text), both with AS(10e,10o) and 500 SPC: comparison between complex overlaps $\langle\psi\vert\psi(t)\rangle$ including hardware experiments at different time steps for the state vector before (SV, full blue circle), after ansatz recompilation (Compiled SV, full green triangle) and measured on a hardware device (Measured, red empty squares). Point mismatch: i)  between state vector calculations before and after recompilation represents divergence due to approximation in the recompilation procedure; ii) between recompilation and hardware experiments represents divergence due to noise in the quantum measurements. Values of the standard deviations over the hardware measured overlaps due to shots sampling are shown in Tabs.~\ref{tab:stderror_rel} and~\ref{tab:stderror_img}
    }
    \label{fig:clusters_overlaps}
\end{figure*}

For considerations on the quality of the ansatz recompilation process and noise deviations due to the quantum device, please compare the data shown in:
1) Figs.~\ref{fig:pu2o3_overlaps_1_1} -- \ref{fig:pu2o3_overlaps_2} for Pu\textsubscript{2}O\textsubscript{3}; 
2) Fig.~\ref{fig:puh2_puh3_overlaps} for PuH\textsubscript{2} and PuH\textsubscript{3}; 
3) Fig.~\ref{fig:clusters_overlaps} for the clusters.

The standard error committed in sampling over 100 or 500 SPC (also referred to as number of shots $N_{\text{SPC}}$)  the complex overlaps on the quantum device are presented in Tabs.~\ref{tab:stderror_rel} and~\ref{tab:stderror_img} 
and calculated as $\sqrt{\frac{1-(\mathfrak{R}\bra{\psi}\ket{\psi(t)})^2}{N_{\text{SPC}}}}$ and similarly for the imaginary part, where $\bra{\psi}\ket{\psi(t)}$ is the complex overlap at time $t$.

\clearpage

\begin{sidewaystable}
    \centering
    \small
    
   \begin{tabular}{ccccccccc}
   \hline

time step & \begin{tabular}[c]{@{}l@{}}Pu\textsubscript{2}O\textsubscript{3} \\AS(2,2) 100 \end{tabular} & \begin{tabular}[c]{@{}l@{}}Pu\textsubscript{2}O\textsubscript{3} \\AS(2,6) 100 \end{tabular} & \begin{tabular}[c]{@{}l@{}}Pu\textsubscript{2}O\textsubscript{3} \\AS(2,8) 100 \end{tabular} & \begin{tabular}[c]{@{}l@{}}Pu\textsubscript{2}O\textsubscript{3} \\AS(2,8) 500 \end{tabular} & \begin{tabular}[c]{@{}l@{}}PuH\textsubscript{2} \\AS(6,8) 500 \end{tabular} & \begin{tabular}[c]{@{}l@{}}PuH\textsubscript{3} \\AS(5,7) 500 \end{tabular}&\begin{tabular}[c]{@{}l@{}}PuH\textsubscript{2}O\textsubscript{2} \\AS(10,10) 500 \end{tabular}&\begin{tabular}[c]{@{}l@{}}PuH\textsubscript{2}2O \\AS(10,10) 500 \end{tabular}\\

\hline
0 & 0.020 & 0.020 & 0.034 & 0.014 & 0.011 & 0.016 & 0.014 & 0.045 \\
1 & 0.020 & 0.078 & 0.081 & 0.039 & 0.029 & 0.023 & 0.043 & 0.045 \\
2 & 0.073 & 0.099 & 0.095 & 0.038 & 0.041 & 0.042 & 0.023 & 0.045 \\
3 & 0.094 & 0.051 & 0.034 & 0.018 & 0.044 & 0.045 & 0.037 & 0.044 \\
4 & 0.099 & 0.057 & 0.085 & 0.042 & 0.028 & 0.038 & 0.033 & 0.028 \\
5 & 0.095 & 0.095 & 0.096 & 0.042 & 0.013 & 0.012 & 0.031 & 0.044 \\
6 & 0.065 & 0.083 & 0.039 & 0.016 & 0.027 & 0.019 & 0.045 & 0.018 \\
7 & 0.020 & 0.034 & 0.083 & 0.036 & 0.043 & 0.033 & 0.045 & 0.043 \\
8 & 0.020 & 0.057 & 0.098 & 0.042 & 0.044 & 0.044 & 0.045 & 0.033 \\
9 & 0.034 & 0.100 & 0.020 & 0.019 & 0.029 & 0.039 & 0.018 & 0.045 \\
10 & 0.075 & 0.073 & 0.089 & 0.031 & 0.011 & 0.025 & 0.043 & 0.019 \\
11 & 0.097 & 0.039 & 0.100 & 0.044 & 0.025 & 0.016 & 0.045 & 0.041 \\
12 & 0.099 & 0.083 & 0.028 & 0.021 & 0.040 & 0.029 & 0.037 & 0.029 \\
13 & 0.091 & 0.092 & 0.067 & 0.029 & 0.044 & 0.042 & 0.039 & 0.045 \\
14 & 0.069 & 0.028 & 0.100 & 0.044 & 0.032 & 0.044 & 0.025 & 0.024 \\
15 & 0.039 & 0.044 & 0.051 & 0.027 & 0.014 & 0.032 & 0.044 & 0.037 \\
16 & 0.028 & 0.098 & 0.051 & 0.033 & 0.022 & 0.015 & 0.031 & 0.030 \\
17 & 0.044 & 0.080 & 0.099 & 0.045 & 0.042 & 0.017 & 0.045 & 0.039 \\
18 & 0.075 & 0.034 & 0.054 & 0.025 & 0.044 & 0.036 & 0.020 & 0.029 \\
19 & 0.098 & 0.054 & 0.071 & 0.032 & 0.035 & 0.045 & 0.039 & 0.031 \\
20 & 0.100 & 0.099 & 0.099 & 0.045 & 0.021 & 0.039 & 0.045 & 0.036 \\
21 & 0.092 & 0.069 & 0.054 & 0.026 & 0.020 & 0.024 & 0.045 & 0.038 \\
22 & 0.060 & 0.020 & 0.054 & 0.026 & 0.040 & 0.015 & 0.041 & 0.033 \\
23 & 0.028 & 0.092 & 0.100 & 0.045 & 0.045 & 0.031 & 0.022 & 0.029 \\
24 & 0.020 & 0.098 & 0.044 & 0.030 & 0.035 & 0.042 & 0.044 & 0.038 \\
25 & 0.065 & 0.047 & 0.051 & 0.019 & 0.015 & 0.044 & 0.045 & 0.040 \\
26 & 0.065 & 0.057 & 0.098 & 0.044 & 0.019 & 0.032 & 0.043 & 0.041 \\
27 & 0.098 & 0.099 & 0.075 & 0.034 & 0.038 & 0.016 & 0.021 & 0.028 \\
28 & 0.099 & 0.090 & 0.047 & 0.029 & 0.045 & 0.019 & 0.037 & 0.042 \\
29 & 0.090 & 0.020 & 0.091 & 0.043 & 0.036 & 0.040 & 0.036 & 0.027 \\
30 & 0.054 & 0.073 & 0.084 & 0.032 & 0.018 & 0.045 & 0.045 & 0.042 \\
31 & 0.034 & 0.100 & 0.047 & 0.017 & 0.021 & 0.041 & 0.042 & 0.021 \\
32 & 0.000 & 0.047 & 0.091 & 0.042 & 0.039 & 0.022 & 0.021 & 0.043 \\
\hline
    \end{tabular}
    \vspace{7pt}
    \caption{Standard deviation for the real part of the overlap ($\mathfrak{R}\langle{\psi\vert\psi(t)}\rangle$) measured on the quantum device and calculated over 100 or 500 SPC. \label{tab:stderror_rel}}

\end{sidewaystable}

\begin{sidewaystable}
    \centering
    \small
    
   \begin{tabular}{ccccccccc}
   \hline
   time step & \begin{tabular}[c]{@{}l@{}}Pu\textsubscript{2}O\textsubscript{3} \\AS(2,2) 100 \end{tabular} & \begin{tabular}[c]{@{}l@{}}Pu\textsubscript{2}O\textsubscript{3} \\AS(2,6) 100 \end{tabular} & \begin{tabular}[c]{@{}l@{}}Pu\textsubscript{2}O\textsubscript{3} \\AS(2,8) 100 \end{tabular} & \begin{tabular}[c]{@{}l@{}}Pu\textsubscript{2}O\textsubscript{3} \\AS(2,8) 500 \end{tabular} & \begin{tabular}[c]{@{}l@{}}PuH\textsubscript{2} \\AS(6,8) 500 \end{tabular} & \begin{tabular}[c]{@{}l@{}}PuH\textsubscript{3} \\AS(5,7) 500 \end{tabular}&\begin{tabular}[c]{@{}l@{}}PuH\textsubscript{2}O\textsubscript{2} \\AS(10,10) 500 \end{tabular}&\begin{tabular}[c]{@{}l@{}}PuH\textsubscript{2}2O \\AS(10,10) 500 \end{tabular}\\
   \hline
    
    0 & 0.100 & 0.100 & 0.099 & 0.045 & 0.045 & 0.045 & 0.045 & 0.045 \\
    1 & 0.097 & 0.057 & 0.077 & 0.027 & 0.038 & 0.038 & 0.018 & 0.044 \\
2 & 0.069 & 0.028 & 0.051 & 0.030 & 0.020 & 0.025 & 0.041 & 0.045 \\
3 & 0.039 & 0.080 & 0.099 & 0.045 & 0.015 & 0.017 & 0.027 & 0.016 \\
4 & 0.020 & 0.089 & 0.057 & 0.028 & 0.036 & 0.031 & 0.036 & 0.044 \\
5 & 0.047 & 0.039 & 0.057 & 0.016 & 0.045 & 0.044 & 0.039 & 0.034 \\
6 & 0.075 & 0.054 & 0.098 & 0.044 & 0.037 & 0.043 & 0.045 & 0.044 \\
7 & 0.091 & 0.096 & 0.069 & 0.031 & 0.018 & 0.031 & 0.045 & 0.019 \\
8 & 0.098 & 0.083 & 0.060 & 0.025 & 0.018 & 0.013 & 0.014 & 0.045 \\
9 & 0.092 & 0.020 & 0.097 & 0.043 & 0.035 & 0.023 & 0.044 & 0.045 \\
10 & 0.063 & 0.063 & 0.078 & 0.037 & 0.044 & 0.039 & 0.019 & 0.042 \\
11 & 0.034 & 0.100 & 0.044 & 0.019 & 0.038 & 0.044 & 0.045 & 0.019 \\
12 & 0.020 & 0.044 & 0.098 & 0.042 & 0.023 & 0.040 & 0.034 & 0.042 \\
13 & 0.060 & 0.020 & 0.081 & 0.036 & 0.014 & 0.023 & 0.033 & 0.045 \\
14 & 0.069 & 0.091 & 0.039 & 0.023 & 0.032 & 0.017 & 0.041 & 0.040 \\
15 & 0.094 & 0.084 & 0.098 & 0.043 & 0.044 & 0.034 & 0.031 & 0.027 \\
16 & 0.100 & 0.039 & 0.083 & 0.037 & 0.039 & 0.044 & 0.045 & 0.037 \\
17 & 0.085 & 0.057 & 0.034 & 0.017 & 0.026 & 0.043 & 0.018 & 0.033 \\
18 & 0.060 & 0.100 & 0.097 & 0.040 & 0.014 & 0.034 & 0.043 & 0.036 \\
19 & 0.028 & 0.078 & 0.077 & 0.037 & 0.030 & 0.016 & 0.026 & 0.034 \\
20 & 0.028 & 0.020 & 0.047 & 0.014 & 0.044 & 0.026 & 0.045 & 0.031 \\
21 & 0.039 & 0.093 & 0.095 & 0.039 & 0.039 & 0.039 & 0.045 & 0.039 \\
22 & 0.073 & 0.098 & 0.092 & 0.042 & 0.026 & 0.044 & 0.029 & 0.033 \\
23 & 0.097 & 0.063 & 0.047 & 0.013 & 0.012 & 0.037 & 0.042 & 0.038 \\
24 & 0.100 & 0.047 & 0.089 & 0.038 & 0.028 & 0.022 & 0.016 & 0.028 \\
25 & 0.092 & 0.097 & 0.093 & 0.041 & 0.043 & 0.015 & 0.045 & 0.044 \\
26 & 0.067 & 0.092 & 0.020 & 0.019 & 0.041 & 0.036 & 0.028 & 0.035 \\
27 & 0.020 & 0.028 & 0.083 & 0.039 & 0.027 & 0.044 & 0.041 & 0.038 \\
28 & 0.028 & 0.075 & 0.094 & 0.043 & 0.015 & 0.040 & 0.026 & 0.033 \\
29 & 0.054 & 0.099 & 0.054 & 0.020 & 0.026 & 0.026 & 0.035 & 0.038 \\
30 & 0.081 & 0.069 & 0.081 & 0.032 & 0.042 & 0.014 & 0.045 & 0.023 \\
31 & 0.085 & 0.020 & 0.098 & 0.043 & 0.044 & 0.028 & 0.025 & 0.042 \\
32 & 0.100 & 0.084 & 0.034 & 0.017 & 0.028 & 0.040 & 0.044 & 0.022 \\
\hline
    \end{tabular}
    \vspace{7pt}
    \caption{Standard deviation for the imaginary part of the overlap ($\mathfrak{I}\langle{\psi\vert\psi(t)}\rangle$) measured on the quantum device and calculated over 100 or 500 SPC. \label{tab:stderror_img}}
\end{sidewaystable}

\clearpage